\documentclass[11pt]{article}
\usepackage[utf8]{inputenc}
\usepackage{fullpage}
\usepackage{authblk}
\usepackage{url}            
\usepackage{booktabs}       
\usepackage{amsfonts, amsthm}       
\usepackage{enumerate}
\usepackage{enumitem}
\usepackage{nicefrac}       
\usepackage{microtype}      
\usepackage{xcolor}         
\usepackage{algorithm, float, setspace}
\usepackage{multirow}
\usepackage[noend]{algpseudocode}
\usepackage{subcaption}
\usepackage{wrapfig}
\usepackage[labelformat=simple]{subcaption}

\definecolor{darkblue}{rgb}{0.0, 0.0, 0.55}
\usepackage[colorlinks=true, citecolor=urlcolor, allcolors=darkblue]{hyperref}

\usepackage[square]{natbib}
\usepackage{mathtools}
\usepackage{comment}

\usepackage{xcolor}

\newcount\Comments  
\newcommand{\kibitz}[2]{\ifnum\Comments=1{\color{#1}{#2}}\fi}

\usepackage{color}
\definecolor{english}{rgb}{0.0, 0.5, 0.0}

\newcount\CommentsA  
\newcommand{\kibitzA}[2]{\ifnum\CommentsA=1{\color{#1}{#2}}\fi}

\newcommand{\wo}{\backslash}

\newcommand{\Algo}[1]{Algorithm~\ref{algo:#1}}

\newcommand{\B}{\mathcal{B}}
\renewcommand{\S}{\mathcal{S}}
\newcommand{\V}{\mathcal{V}}
\newcommand{\N}{\mathcal{N}}
\newcommand{\vr}{\mathbf{r}}

\newcommand{\X}{\mathcal{X}}
\newcommand{\A}{\mathcal{A}}
\newcommand{\R}{\mathcal{R}}
\renewcommand{\P}{\mathcal{P}}
\renewcommand{\O}{\mathcal{O}}

\newcommand{\I}{\mathcal{I}}

\newcommand{\ind}{\mathbb{I}}

\newcommand{\vtheta}{\boldsymbol{\theta}}

\newcommand{\E}{\mathbb{E}}

\newcommand{\card}[1]{\lvert#1\rvert}

\newcommand{\Parens}[1]{\left(#1\right)}

\newcommand{\norm}[1]{\lVert#1\rVert}

\DeclareMathOperator*{\argmax}{argmax}

\begin{document}

\title{Platform Behavior under Market Shocks: \\A Simulation Framework and Reinforcement-Learning Based Study%
\thanks{This research is funded in part by Defense Advanced Research Projects Agency under Cooperative Agreement HR00111920029. The content of the information does not necessarily reflect the position or the policy of the Government, and no official endorsement should be inferred. This is approved for public release; distribution is unlimited.\looseness=-1}
}

\author[1]{Xintong Wang}
\author[1]{Gary Qiurui Ma}
\author[1]{Alon Eden}
\author[1]{Clara Li}
\author[2]{Alexander Trott}
\author[2]{Stephan Zheng}
\author[1]{David C. Parkes}
\affil[1]{Harvard University}
\affil[1]{\texttt {\{xintongw,qiurui\_ma,aloneden\}@seas.harvard.edu, clarali@college.harvard.edu, parkes@eecs.harvard.edu}}
\affil[2]{Salesforce Research}
\affil[2]{\texttt {\{atrott,stephan.zheng\}@salesforce.com}}

\date{\vspace{-6ex}}

\maketitle
\begin{abstract}
    We study the behavior of an economic platform (e.g., Amazon, Uber Eats, Instacart) under shocks, such as COVID-19 lockdowns, and the effect of different regulation considerations imposed on a platform. 
    To this end, we develop a multi-agent Gym environment of a platform economy in a dynamic, multi-period setting, with the possible occurrence of economic shocks.
    Buyers and sellers are modeled as economically-motivated agents, choosing whether or not to pay corresponding fees to use the platform.
    We formulate the platform's problem as a \emph{partially observable Markov decision process}, and use deep reinforcement learning to model its fee setting and matching behavior. 
    We consider two major types of regulation frameworks: (1)~taxation policies and (2)~platform fee restrictions, and offer extensive simulated experiments to characterize regulatory tradeoffs under optimal platform responses.
    Our results show that while many interventions are ineffective with a sophisticated platform actor, we identify a particular kind of regulation---fixing fees to optimal, pre-shock fees while still allowing a platform to choose how to match buyer demands to sellers---as promoting the efficiency, seller diversity, and resilience of the overall economic system.
    %
    %
    %
\end{abstract}

\section{Introduction}
Market-driven platforms, such as  Amazon, DoorDash, Uber, and TaskRabbit, play an increasingly important role in today’s economy, bringing together parties  to facilitate trade and presenting new ways to create value.
First, they reduce \textit{search cost} by introducing potential matches that were not known before, and second, they reduce \textit{fulfillment cost} by taking care of service or product delivery, thus reducing the effort made to complete transactions.
%
The importance of the platform-based economy became even more stark during the COVID-19 pandemic, and especially for the restaurant industry.
Stay-at-home orders, together with the closure of dine-in channels and caution in regard to visiting brick-and-mortar businesses,  tremendously increased the fulfillment cost of consumers transacting in physical locations. 
This promoted the role of platforms, for example leading
 an increasing number of users and restaurants  to adopt food-delivery platforms. One
study of Uber Eats from February through May 2020  showed a surge in both demand and supply after the shelter-in-place guidance was issued in several states in the U.S.~\citep{raj2021}.

At the same time,  this new prominence has provided platforms with increased  market power. 
As a demonstration, restaurant commission fees are being 
set by some platforms to as high as 30\% per order, leaving traditional restaurants, many of whom  may no longer have dine-in revenue, with low or even negative margins~\citep{mckinsey2021}. 
In December 2020, the National Restaurant Association reported that more than 110,000 U.S.~restaurants---one in six---have permanently closed down since the start of pandemic~\citep{NRA2020}.
To support restaurants, many states such as New York and California have imposed commission caps, and yet platforms have responded with countermeasures. As an example, the day after Jersey City enforced a 10\% cap on fees, Uber Eats added a \$3 delivery fee to customers and reduced the delivery radius for restaurants.%
\footnote{https://www.protocol.com/delivery-commission-caps-uber-eats-grubhub}
This speaks to the complexity of the ecosystem, and regulatory policies on third-party delivery platforms have been continuously proposed and debated as the pandemic calms down.

%
%
%
In this work, we use the methods of AI to study a platform-based economy under market shocks through a multi-agent simulation framework, with this as a first step towards reproducing phenomena observed in the real-world platform economy and as a tool for conducting counterfactual analysis to understand platform behavior in response to different regulations. %
We develop a multi-agent Gym environment to capture major aspects of a platform economy in a multi-period setting, with key modeling choices based on the existing economic literature (e.g., epoch-based decision making, user behavior inertia, and the discrete-logit choice model).
Our model captures a full cycle of market shock,
 designed to represent the pre-, during, and post-crisis periods.

We formulate the platform's problem as a \emph{partially observable Markov decision process} (POMDP), with both commonly observable components (e.g., shocks) and private elements (e.g., buyers' knowledge about sellers and transactions made off the platform),
and model the platform as a rational agent that uses {\em reinforcement learning} (RL) to set fees and match buyer queries to sellers. 
Buyers and sellers decide  whether or not to join the platform, considering both fees and their search and fulfillment costs; 
on-platform buyers further decide their preferred channel to make transactions, i.e., either with a recommended platform seller or with a known seller off the platform.
We conduct extensive simulationed experiments to explore a range of settings, these differing in market structures (i.e.,  locations of buyers and sellers in the product and preference space), the knowledge levels of buyers about sellers, and the cost of off-platform fulfillment. We demonstrate how to
successfully use RL to model the optimal behavior of a platform under different regulatory considerations, and study the effect of a platform on the efficiency and resilience of the overall economic system. 

In the absence of any regulation, we find that a revenue-maximizing platform, even while helping to facilitate trades during the market shock, tends to leverage its increased market power during the 
shock to raise fees and extract surplus from buyers and sellers,  leading to seller shutdowns and lower post-shock economic welfare when compared with the pre-shock economic welfare.

As a first kind of regulatory intervention, we consider a class of {\em taxation policies} that impose tax rates on different categories of profit (e.g., revenue made from user subscriptions or revenue made from transaction referrals), and study the platform response and its effect on the economy. 
We further study regulations that \textit{cap platform fees} in a particular way.
Our first main result shows that either taxation or capping to some subset of fees simply leads the platform to transfer incremental costs to  users by adjusting other fees,  demonstrating the power of an autonomous, strategic platform agent that is represented by the learned RL policies in modeling rational behavior.
On the other hand,  we show that capping all fees (i.e., not just a subset)
 has a moderately positive effect on protecting sellers from bankruptcy and promoting a resilient ecosystem. 
In practice, this requires detailed domain knowledge on the part of a regulator in regard to how to effectively set these caps.
Another limitation is that, for reasons of simulation tractability and interpretability, we only study this intervention
 in the presence of a platform that follows a myopically optimal matching policy.

In this light, our second main result is especially interesting, as in these settings, we study a platform that is required
to keep the same fee structure that it chooses to use in the absence of shocks 
but has the flexibility in choosing how to match buyer demands to sellers.  Thus, it  requires 
no special knowledge on the part of a regulator and still gives the platform full flexibility in regard to parts of its behavior that are usually proprietary and not easily regulatable. 
We show that a revenue-maximizing platform learns, under this intervention, 
to use matching to retain a more diverse set of sellers on platform, so as to increase its long-term revenue from user registrations, while also helping to promote the efficiency, resilience, and seller diversity of the overall economy.
%
%
%

We aim that the present framework, which combines a multi-agent Gym environment with the use of RL to derive optimal platform responses, provides a tool to use in understanding regulations and platform economies, complementing economic theory that often becomes analytically intractable in complex and dynamic environments and pure data-driven approaches that cannot answer questions about changing market and agent designs.
We return to a critical discussion of the opportunities and outstanding challenges with this kind of AI-based approach at the end of the paper.

\subsection{Related Work}
\paragraph{Platform Models.} 
There is an extensive economic literature that focuses on how to establish network effects through platform fee-setting or subsidizing one side of the market under various forms of platform competition (e.g., single- vs. multi-homing)~\citep{Caillaud2003, Rochet2003, Armstrong2006}.
%
%
\cite{Caillaud2003} study competition among platforms that act as matchmakers for homogeneous buyers and sellers and use both registration and transaction fees.
In a two-stage model where platforms first set fees and then users simultaneously choose which platforms to register, the authors show that dominant firms are better off charging transactions rather than registrations and that competition is more intense when platforms cannot deter multi-homing.
\cite{Rochet2003} generalize the setting to heterogeneous users and analyze a model in the context of credit card market with charges imposed only on a per-transaction basis.
They show that the volume of transactions and the profit of a platform depend not only on the total fees charged to the parties to the transaction, but also on its decomposition.
\cite{Armstrong2006} further investigates such fee-setting behavior by varying the magnitude of \textit{cross-group externalities}, through considering a monopoly platform, a model of competing platforms with single-homing agents, and a model of ``competitive bottlenecks'' with one group of users joining all platforms.
To facilitate equilibrium analysis, these models often rely on simplified assumptions for tractability, e.g., restricting to a single round of platform fee-setting and agent subscribing, adopting a fixed platform matching policy, and assuming homogeneous non-platform actors.

We extend economic models of the above pioneering articles to multi-period, dynamic settings for studying platform behavior under market shocks, where (1)~a shock may occur to change agents' price elasticity of demand in using the platform, (2)~agents can flexibly join or quit the platform in response to shocks and fees, and (3)~transactions can be completed via the platform or in the brick-and-mortar traditional market.
Many of these features are essential to our problem at hand, and may not be removed or easily stylized for tractability. 
For such reasons, we consider a platform that uses RL to set fees and adjust matching strategies dynamically.

%
%
%
Besides related literature that characterizes platform behavior, there are also works on complexity results in regard to setting optimal fees~\citep{PapadimitriouPP16}, optimally matching buy queries to sellers~\citep{Mladenov2020}, and designing a platform for each agent state~\citep{pdp}.
\citet{PapadimitriouPP16} show that setting optimal fees by a revenue-maximizing mechanism is NP-hard, even for a market with one buyer and two periods.
Optimally matching buy queries to sellers has also been proved computationally hard for a single-period setting with all non-platform actors fixed and no platform fees~\citep{Mladenov2020}.
In a different setting, \citet{pdp} explore the design of a platform for each of the agent's activities or states, under which the agent's transition probabilities of the Markov chain are affected and the designer aims to maximize engagement.
It is proven that the designer's optimization problem is NP-hard to approximate within any finite ratio.
Our setting differs in that we focus on a market-based platform with network effects among agents (buyers and sellers).
%

Existing literature on the role and behavior of economic platforms under market shocks is fairly limited.
Empirical studies have been conducted to characterize the impact of COVID-19 on the supply and demand of food-delivery platforms~\citep{raj2021}, the effect of commission fee caps on on-demand services~\citep{li2021}, as well as the extent to which pandemic has persistently changed consumer's purchasing behavior, even after the shock calms down due to habit formation around delivery~\citep{oblander2022}. 
To the best of our knowledge, we are not aware of any theoretical or simulation studies that model or investigate the strategic behavior of platform in response to market shocks and various forms of regulations.


\paragraph{Using Reinforcement Learning for Economic System Design.}
There has been considerable recent interest in making use of methods of AI, and especially RL, for understanding
the design of complex economic systems,
though we are not aware of previous work on the use of RL or AI for the modeling of market-based platforms.
Existing work on the use of RL for economic system design
includes the problem of  dynamically setting reserve prices in auctions~\citep{ShenPLZQHGDLT20}, learning how to sell user impressions to advertisers~\citep{Tang17abc}, the design of tax policies in multi-period economies~\citep{Zheng2021},  optimizing user satisfaction for multi-stage recommender systems~\citep{Chen2019,Zhan2021},  and the design of sequential price mechanisms~\citep{BreroEGPR21}.\looseness=-1

Many of these works make use of agent-based simulations to model actors in the economic system and conduct counterfactual analysis through the dynamic interactions of agents. 
For example,~\citet{Zheng2021} use RL to model a social planner who designs income taxes in multi-period, simulated spatial economies. 
Similar to ours is the work of \citet{Zhan2021}, which is built on the {\em RecSim environment}~\citep{Ie2019} and adopts RL to optimize the long-term social welfare of both users and content providers in a dynamic recommender system. 
Besides the presence of economic shocks, our setting differs in the use of platform fees, the existence of an alternate sales channel (i.e., off-platform transactions), and the implication that agents can choose to join or quit the platform.

\section{Motivating Example: A Simple Platform Economy} 
\label{sec:example}

Before introducing the full model, we use a simple economic system under market shocks to illustrate that a revenue-maximizing platform with typical myopic matching policy may be suboptimal from the viewpoint of overall economic efficiency, while a platform that (i)~uses more sophisticated matching of buyer demands to sellers or (ii)~is surplus-aware in optimizing for a combination of its own revenue and the on-platform buyer and seller surplus may help to improve market outcomes.

\begin{figure}[t]
	\centering
	\includegraphics[trim={0cm 0cm 0cm 0},width=0.9\columnwidth]{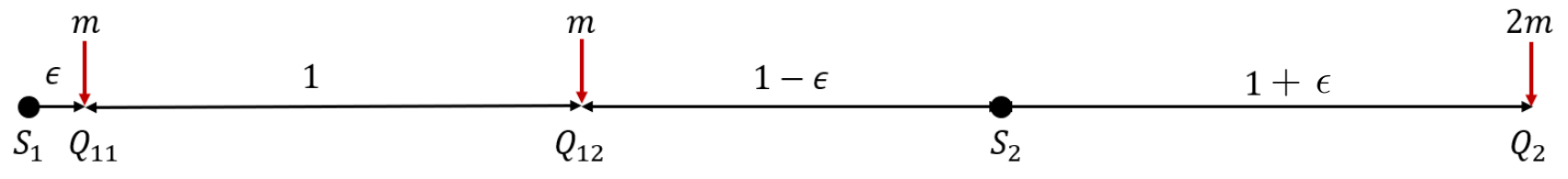}
	\caption{A simple platform economy with two buyers and two sellers located in
 a one-dimensional latent structure. Buyer 1 has $m$ queries at each of the locations $Q_{11}$ and $Q_{12}$
 and knows Seller 2.   Buyer 2  has $2m$ queries at location $Q_2$ and knows both sellers.
	\label{fig:motivating_example} \label{fig:basic_setup}}
\end{figure}

We model an economy with two buyers and two sellers, where the sellers differ in the kinds of product they offer and the buyers differ in the kinds of product they are interested in.
In particular, we represent the features of sellers and buyer queries in a one-dimensional \textit{latent space} on the real line from $[-2,1+\epsilon]$, treating $S_2$ as the origin (see Figure~\ref{fig:basic_setup}). 
Each buyer generates demand
for different kinds of products through submitting a {\em query}, which is a statement of interest for a product with a particular feature.
When  matched with a seller $s$, a query $q$ gives
 the buyer a \textit{matching utility} of $u(q,s)$, with this value being inversely proportional to the distance between $q$ and $s$ to reflect matching quality. 
%
%
%
Each seller is always 
willing to complete transactions with buyers off-platform, but this is
only possible with a buyer who 
has knowledge of the seller. 
In this example, we assume that Buyer 1 only knows Seller 2, while  
Buyer 2 knows both Seller 1 and Seller 2.
The effect is that Buyer 1 can only transact with Seller 1 if both of them are on the platform. 

Buyers and sellers make decisions at the start of each \textit{epoch} (e.g., a month) about whether or not to join the platform, incurring an associated registration fee in the case of joining.  
 Within an epoch, we model each 
buyer as issuing $2m$ queries, for $m>0$, 
with these queries distributed
as shown in Figure~\ref{fig:basic_setup}. 
Buyer 1 issues $m$ queries at location $Q_{11}$ (with a distance of $\epsilon$ from Seller 1
 and $2-\epsilon$ from Seller 2) and $m$ queries at location $Q_{12}$
 (with a distance of $1+\epsilon$ from Seller 1 and $1-\epsilon$ from Seller 2).  
 Buyer 2 issues  $2m$ queries  at location $Q_2$
 (with a distance of  $3+\epsilon$ from Seller 1 and $1+\epsilon$ from Seller 2).\looseness=-1

The platform sets a {\em per-epoch, buyer registration fee}, $P_\B\geq 0$, and a
{\em per-epoch, seller registration fee}, $P_\S$, and recommends an on-platform seller for each query from a buyer that is on the platform.%
\footnote{In the full model, we further include a per-transaction seller referral rate that can be set and charged by a platform.}
A buyer can choose among the options of (i)~transacting with a known seller off-platform, (ii) if on platform, transacting with the platform-recommended seller, and (3) not to transact with any seller.

Besides the benefit of being matched to a new seller, the platform also reduces the fulfillment cost
when a transaction is completed on platform. 
We model this through the \textit{world transaction friction}, $\mu>0$, capturing the effort required by a buyer to transact off platform, such as visiting a restaurant.
Therefore, the surplus for a buyer with query $q$ that transacts with seller $s$ is
$r(q,s,\I) = u(q,s)-\mu \cdot \I_w,$ 
where  $\I_w \in \{0,1\}$ indicates whether a transaction is completed off-platform ($\I_w=1$) or via the platform $(\I_w=0)$.
%
%
%
Each seller is modeled as receiving a surplus of 1 from each completed transaction, regardless of the matched buyer, and has a {\em fixed cost} of $m$ per epoch.
A seller whose cost is not covered by surplus received from serving buyer queries will shut down and can no longer be matched to future queries.  

We study this simple economy in an environment where
the matching utility function is $u(q, s) = 2 - |q-s|$, with a high and constant world transaction friction of $\mu=1$.
This can be considered to model the COVID pandemic lockdowns.%
\footnote{In our full model, we consider a setting where there is a pre-, during- and post-shock period, but here we in effect study only the during-shock period.}
Beyond setting fees,
the platform also decides how it will match any queries executed on platform with sellers. 
One typical choice is myopic matching, where the platform always
 matches a query $q$ to the seller who gives the highest matching utility, i.e., $s_q^* = \argmax_{s \in \S} u(q,s)$,
 considering the set $\S$ of on-platform sellers. More generally, in the case of having both sellers
 on the platform, the platform
 can decide which fraction of queries in which locations to match to each seller.

In studying the simple economy, 
we compute the {\em Stackelberg equilibrium} where the platform makes decisions, in regard to registration fees and perhaps also matching policies, for each epoch.
Buyers and sellers then best respond and form an equilibrium,
 where the decision of a buyer or seller
 is whether or not to join the platform (and subsequently how to transact on the basis of
 off-platform and on-platform opportunities).
The policy of a platform forms a Stackelberg equilibrium when the platform cannot increase its objective value by choosing to commit to another policy. 

In what follows, we consider the baseline of the absence of a platform, as well as three different cases,
which vary in regard to 
the matching policy of the platform and whether or not 
the surplus to buyers and sellers is part of the platform's objective (in addition to its own revenue). 
We provide a high-level description of each case and summarize key results in Table~\ref{table:motivating_example}. 
Complete proofs of the equilibrium
structure are deferred to Appendix~\ref{sec:apx-example}.
%
%
\begin{itemize}[leftmargin=*]
    \item \textbf{Case 1: No platform.}
    In the absence of a platform, the market is very inefficient under shocks: transactions happened only between the $Q_{12}$-type queries 
    from Buyer 2 and Seller 2.
    Seller 1 gets no transaction and shuts down.
    The per-epoch welfare of the system (i.e., the sum of buyer and seller surplus) is $\epsilon m$.\looseness=-1
    %
    %
    \item \textbf{Case 2: A revenue-maximizing platform with myopic matching.}
    This case considers a platform making an equilibrium choice about registration fees, while fixing the matching policy to be myopic optimal. 
In effect, the platform chooses which of the buyers and sellers to bring to the platform, setting
fees to maximize revenue while keeping these agents on the platform.
    In the Stackelberg equilibrium, the fees are set to $P_\B=(1+\epsilon)m$ and $P_\S=3m$, and the two buyers and Seller 2 join the platform, with all queries directed to Seller 2.
    The effect is that Buyer 1 gets a per-epoch surplus of $\epsilon m$ ($=r(Q_{11},S_2, 0) \cdot m + r(Q_{12},S_2, 0) \cdot m - P_\B$), and Buyer 2 gets a per-epoch surplus of $(1-3\epsilon)m$ ($=r(Q_{2},S_2, 0) \cdot 2m - P_\B$).
    Since $P_\S$ is set to $3m$ and each seller has a fixed cost of $m$, Seller 1 stays off-platform and goes bankrupt, whereas Seller 2 is matched to $4m$ queries but receives zero surplus.
    The platform's per-epoch
revenue is $(5+2\epsilon)m$ ($=2P_\B + P_\S$)
 and the per-epoch economic welfare (i.e., the sum of platform revenue and buyer and seller surplus) is $6m$, which is much larger than the case without a platform.
    \item \textbf{Case 3: A revenue-maximizing platform that rationally sets fees and matching policy.}
    This case  allows the platform to both set fees and choose
    to deviate from a myopic matching policy, for example
directing more queries  than just the $Q_{11}$ queries from Buyer 1 
to Seller 1. 
    In the Stackelberg equilibrium, the platform brings both sellers and both buyers to join the platform, achieving this by setting $P_\B = (2-2\epsilon)m, P_\S=m$, and matching all queries from Buyer 1 to Seller 1 and all queries from Buyer 2 to Seller 2. 
    Buyer 1 gets a per-epoch surplus of $m$ ($=r(Q_{11},S_1, 0) \cdot m + r(Q_{12},S_1, 0) \cdot m - P_\B$), and Buyer 2 gets a per-epoch surplus of $0$ ($=r(Q_{2},S_2, 0) \cdot 2m - P_\B$).
    With $P_\S$ set to $m$, each seller is  matched to $2m$ queries and has zero surplus.
    The platform revenue is $(6-4\epsilon)m$ ($=2P_\B + 2P_\S$), and the total welfare of the system is $(7-4\epsilon)m$.
    In this case, the platform is able to keep both sellers from going bankrupt 
through the use of non-myopic matching, with the effect of
 improving its own revenue as well the system welfare, compared with Case 2 of a revenue-maximizing 
 platform that uses myopic matching.
    \item \textbf{Case 4: A surplus-aware platform with myopic matching.}
   In this case, we consider a platform that is motivated
   to optimize for a combination of its own revenue and the  surplus of on-platform buyers and sellers, i.e., $\text{rev} + \alpha \cdot \text{surplus}$, where $\alpha < 1$ is the weight placed
   on user surplus in the objective. 
   This can be motivated either as a simple model of the effect
   of regulation or the motivations of
 a more socially conscious platform.
When $\alpha > 0.5$, the platform will choose 
fees  $P_\B=(2-2\epsilon)m$ and $P_\S=0$ and all agents join the platform with $Q_{11}$-type queries from Buyer 1 matched to Seller 1 and the rest of queries
to Seller 2.
    Under such fees, Buyer 1 has a per-epoch surplus of $(1+2\epsilon) m$ ($=r(Q_{11},S_1, 0) \cdot m + r(Q_{12},S_2, 0) \cdot m - P_\B$), and Buyer 2 has a per-epoch surplus of 0 ($=r(Q_{2},S_2, 0) \cdot 2m -P_\B$). 
    Seller 1 gets $m$ queries and has  zero surplus whereas Seller 2 gets $3m$ queries and  a per-epoch surplus of $2m$.
    The per-epoch platform revenue is $(4-4\epsilon)m$ ($=2P_\B + 2P_\S$), and the per-epoch 
    overall welfare is $(7-2\epsilon)m$.
    We note that, in this case, all queries are matched to their ideal seller, and the system achieves the maximum possible welfare.
\end{itemize}

\begin{table}[t]
    \centering
    \scriptsize   
    \begin{tabular}{p{0.03\textwidth}p{0.05\textwidth}p{0.1\textwidth}lllll}
        & &\textbf{No platf.} & \textbf{Revenue-maximizing} & \textbf{Rational matching} & \textbf{Surplus-aware ($\alpha$)} & \textbf{Ideal}\\[0.2em]
        \hline
        \midrule
        System &Welfare & $\epsilon m$ & $6m$ & $(7-4\epsilon)m$ &$(7-2\epsilon)m$ &$(7-2\epsilon)m$\\[0.2em]
        \midrule
        \multirow{2}{*}{B1} & Surplus & $\epsilon m$ & $\epsilon m$ & $m$ & $(1+2\epsilon)m$ & $3m$\\[0.2em]
        & Matches & $Q_{12} \rightarrow S_2$ & $Q_{11}, Q_{12} \rightarrow S_2$ & $Q_{11}, Q_{12} \rightarrow S_1$ &$Q_{11} \rightarrow S_1, Q_{12} \rightarrow S_2$ &$Q_{11} \rightarrow S_1, Q_{12} \rightarrow S_2$\\[0.4em]
        \multirow{2}{*}{B2} &Surplus & 0 & $(1-3\epsilon) m$ & 0 & 0 & $(2-2\epsilon)m$\\[0.2em]
        & Matches & None & $Q_2 \rightarrow S_2$ & $Q_2 \rightarrow S_2$ & $Q_2 \rightarrow S_2$ & $Q_2 \rightarrow S_2$\\[0.4em]
        \multirow{2}{*}{S1} & Surplus & 0 & 0 & 0 & 0 & 0\\[0.2em]
        & State & Bankrupt & Bankrupt & $2m$ queries & $m$ queries & $m$ queries\\[0.4em]
        \multirow{2}{*}{S2} & Surplus & 0 & 0 & 0 & $2m$ & $2m$\\[0.2em]
        & State & $m$ queries & $4m$ queries & $2m$ queries & $3m$ queries & $3m$ queries\\[0.4em]
        \multirow{2}{*}{Platf.} & Revenue & 0 & $(5+2\epsilon)m$ & $(6-4\epsilon)m$ & $(4-4\epsilon)m$ & 0\\[0.2em]
        & Fees & N/A & $P_\B=(1+\epsilon)m, P_\S=3m$ & $P_\B=(2-2\epsilon)m, P_\S=m$ & $P_\B=(2-2\epsilon)m, P_\S=0$ & $P_\B=0, P_\S=0$\\[0.2em]
        %
        \hline\hline
    \end{tabular}
    \vspace{1ex}
    \caption{Total system welfare, each buyer or seller agent surplus, and platform revenue achieved in the Stackelberg equilibrium of the simple economy with no platform (Case 1), a revenue-maximizing platform (Case 2), a rational-matching platform (Case 3), a surplus-aware platform (Case 4), and an ideal world where each buyer knows all sellers and there is no fee or transaction friction.
    The second row of the entry for each agent describes buyer query matching,
seller agent states, or fees set by the platform.
    \label{table:motivating_example}}
\end{table}

This simple economy is helpful in demonstrating the potential effectiveness of adopting non-myopic matching  and imposing market regulation (modeled simplistically in this case in the form of adding a surplus term to the platform's objective)
 in improving the efficiency of the ecosystem and also avoiding seller bankruptcy.
Next, we extend to a more general multi-agent platform model, with the addition of a per-transaction referral rate, complex market structures (i.e., latent locations of buyers and sellers) and buyers' knowledge about sellers, agent behavior inertia, changing world transaction friction to represent full cycles of market shock, and different forms of regulations.
Given the complexity and intractability of the dynamic environment, we study a platform that uses RL to respectively learn its fee-setting policy and matching policy.

\section{A Multi-Agent Platform Model}
\label{sec:gym_env}

This section formulates the dynamic, multi-agent model, with which we simulate a platform-based economy. 
We defer details of learning the platform's fee-setting and matching policy to Section~\ref{sec:pdp}. 

\subsection{Market Environment and Agent Dynamics}
\label{sec:dynamics}
Consider a market populated with a set of heterogeneous buyers $\B$,  a set of heterogeneous sellers $\S$, and a single platform agent $p$.
We follow the embedding-based representations in recommender systems~\citep{Salakhutdinov2007} and use a two-dimensional {\em latent space} to represent each buyer or seller agent, i.e., $v_b, v_s \in \V \subseteq [0, 1]^2$. 
The first dimension of an agent's {\em location}, denoted $v^0$, describes product features (e.g., in the case of food, the type of cuisine, Italian, Mediterranean or Japanese, the taste of dishes, spicy or not), and the second dimension, denoted $v^1$, the normalized price level (e.g., $\$\ldots\$\$\$\$$). 
A seller $s$ offers food at a price $v^1_s$, 
of which an $\omega_s$ fraction is the per-unit production cost.
Each buyer $b$ knows a subset of sellers $\S_b \subseteq \S$ and may transact with $s \in \S_b$ without using a platform. 
Buyers who use the platform are also introduced to additional sellers. 
%

We formulate an {\em epoch-based} decision problem for agents, similar to the recommendation problem in~\citet{Mladenov2020}. 
Within an episode, there are multiple epochs, indexed $k$, and each epoch has a fixed length of $T$ time steps (e.g., a month of 30 days). 
We consider a macro-level {\em world transaction friction}, denoted $\mu_k>0$, which varies by epoch and represents the cost of buyers completing a transaction off the platform (see Equation~\eqref{eq:potential_world_matching_surplus}).
We model {\em shocks}, corresponding to changes in this friction. 
For example, during a pandemic, transaction frictions for the food-service industry were extremely high, due to fears of sharing indoor spaces and the absence of dine-in options.  
Below we describe agent dynamics within an epoch.

\vspace{-1ex}
\paragraph{\textbf{At the start of an epoch $k$.}}
The platform sets fees, including the  buyer and seller {\em subscription fees}, denoted $P_{\B, k}\geq 0$ and $P_{\S, k}\geq 0$ respectively, 
and a per-transaction seller {\em referral rate} $P_{R, k}\in [0,1]$, the fraction of prices as transaction fee paid by the seller to the platform. 
We discuss the platform  policy in setting fees in Section~\ref{sec:pricing_policy}.
%

Buyers and sellers  observe the fees and the world transaction friction $\mu_k$,  and decide whether to pay the subscription fee to use the platform for epoch $k$. 
We denote the sets of subscribed buyers and sellers  in epoch $k$ as $\B_k$ and $\S_k$.
We assume that the platform knows the locations of buyer queries and on-platform sellers in the latent space. 
This reflects that platforms tend to have good data on the market-relevant properties of sellers and that the combination of a search interface and historical buyer information gives good information on the current demand context of a buyer.
Each buyer has a {\em per-epoch budget constraint}, denoted $\psi_b>0$, which is linearly proportional to the buyer's price preference $v^1_b$. 
This controls the number of transactions that a buyer will undertake in a given epoch.
%
We discuss how buyers and sellers decide whether or not to join the platform for the upcoming epoch in Section~\ref{sec:choice_subscribe}.

Beyond fees, the platform matches queries from platform buyers to platform sellers, with a matching policy learned through RL
 (see Section~\ref{sec:matching_policy}). 

\begin{figure}[t]
	\centering
	\includegraphics[width=0.95\columnwidth]{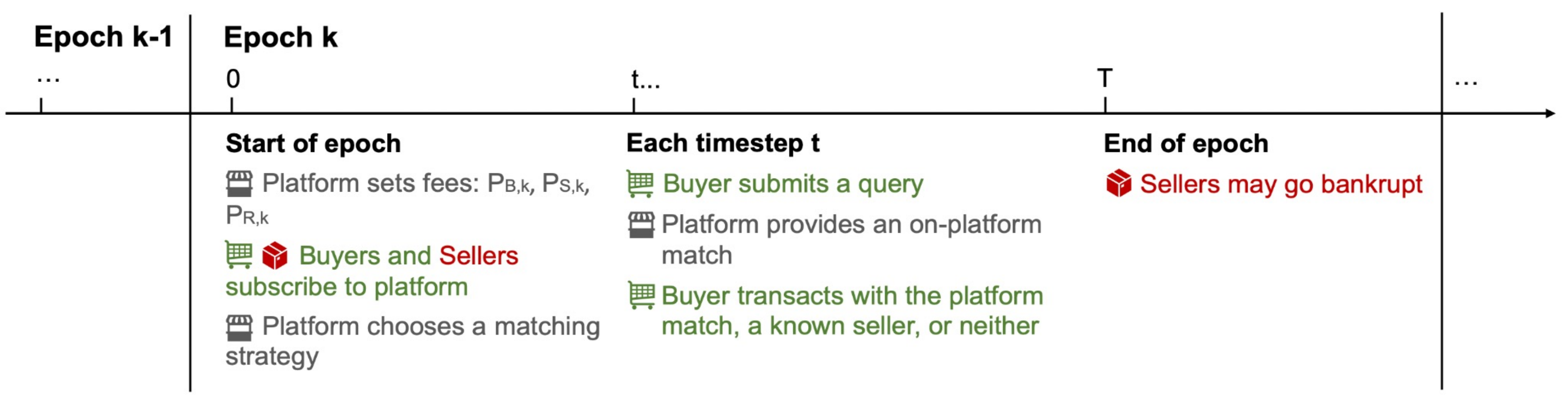}
	\caption{An illustration of the epoch-based decision making for buyers, sellers, and the platform.
	\label{fig:model_timeline}}
\end{figure}

\vspace{-1ex}
\paragraph{\textbf{For each time step $t$ in an epoch $k$.}} 
We follow the sequence of ``query, match, and transact'':

\begin{enumerate}[label=(\roman*)]
    \item \emph{Query}: a buyer $b \in \B$ is randomly selected to submit a query according to their taste and price preferences (e.g., ordering \$\$\$\$ sushi or \$\$ pizza), $q_{b,t} \sim \N(v_b, \sigma^2_b)$, where $\sigma_b$ specifies the query variance of $b$ around their latent location.
    \item \emph{Match}: for an on-platform buyer $b$, the platform observes $q_{b,t}$ and matches it to an on-platform seller, denoted $s_{p,t} \in \S_k$.
    \item \emph{Transact}: the buyer $b$ can pick a seller $s \in \{s_{p, t}\} \bigcup \S_b$ if $b$ is on-platform, and $s \in \S_b$ otherwise. 
    A buyer can also choose not to transact if they are not satisfied with any seller, i.e., the matching surplus is negative (details in Section~\ref{sec:choice_transact}).
\end{enumerate}
We refer to a transaction that is matched via the platform as \emph{a platform transaction}, and otherwise \emph{a world transaction}.
We note that an on-platform buyer may choose to complete a world transaction if the platform suggests an inferior match.
In addition, a buyer may have a world transaction with a seller known to the buyer, whether or not the seller is on the platform, and by such, a seller on the platform is also able to complete off-platform transactions.
For each world transaction, the buyer suffers a fulfillment cost of $\mu_k$,  while for each platform transaction, the seller pays a {\em referral fee}, which is a fraction $P_{R, k}$ of the seller's price.
%

\paragraph{\textbf{At the end of an epoch $k$.}}
Buyers and sellers evaluate their surplus from transactions and fees paid in order to guide future subscription decisions.
The platform evaluates the revenue made through subscription and referral fees and can make adjustments to fees
or its matching policy (see Section~\ref{sec:pdp}).
Each seller has a {\em shutdown threshold}, $\lambda_s \in \mathbb{N}_{> 0}$, and will go bankrupt if they do not obtain enough number of transactions or surplus to meet the threshold. 
Once bankrupt, a seller is unable to engage in transactions in future epochs.


\subsection{Transaction-Level Decisions}
\label{sec:choice_transact}

A buyer with query $q$ who transacts with seller $s$ receives a \emph{matching utility}, $u_{\B}(q,s)$, reflecting the matching quality.
A buyer with a choice of transactions prefers the one that maximizes immediate \emph{matching surplus}, defined as matching utility minus any transaction friction (i.e., the buyer is myopic without considering how their current choice will affect sellers and the availability of future options).

\paragraph{Choice in the world.}
\label{sec:choice_world_only}
A buyer $b$ with query $q_{b, t}$ can choose from  known sellers  whose prices (denoted by $v_s^1$ for seller $s$) are within their epoch-budget left at $t$, denoted $\psi_{b,t}$, i.e.,
$\S_{b, t} := \{s \in \S_b \ :\  v^1_s \leq \psi_{b,t}\}$.
%
For $\S_{b, t} \neq \emptyset$, the best world choice is
$s^*_w := \argmax_{s \in \S_{b,t}} u_{\B}(q_{b,t}, s)$. 
Since the world transaction friction $\mu_k$ could be high enough to prevent a buyer from transacting, we let $s^w_{b, t}$ be $s^*_w$ if $u_{\B}(q_{b,t}, s^*_w) > \mu_k$, and $\phi$ otherwise, 
to denote no preferred world seller.
%
Writing $u_{\B}(q_{b,t}, \phi)=0$, the \emph{world surplus} to buyer $b$ at time $t$  is
\begin{equation}
	\label{eq:potential_world_matching_surplus}
	u^w_{b, t} = 
	\max(u_{\B}(q_{b,t}, s^w_{b, t}) - \mu_k,0).
\end{equation}

\paragraph{Choice on the platform.}
\label{sec:choice_platform_only}
For an on-platform buyer $b$, the \emph{platform surplus}, $u^p_{b, t}$, at time $t$ is $u^p_{b, t}=u_{\B}(q_{b,t}, s_{p, t})$ in the case that the platform-recommended seller, $s_{p, t}$, is within the remaining budget $\psi_{b,t}$, and 0 otherwise.
For an off-platform buyer $b$, we set $u^p_{b, t}=0$.

\paragraph{Overall choice.}
%
If no seller provides a positive surplus, the buyer will choose not to transact. Otherwise, a
 buyer $b$ chooses $s^w_{b, t}$ if it is off-platform, and the most preferred of $s^w_{b, t}$ and $s_{p, t}$ when it is on-platform.%
\footnote{We note that $s_{p, t}$ and $s^w_{b, t}$ may be the same seller, in which case the buyer will choose to transact via the platform since the world transaction friction is always positive.}
We write $s_{b,t}$ to  denote the choice of the buyer at time $t$,
and denote a buyer's query, off- and on-platform seller options, and the chosen seller for transaction as the 4-tuple $(q_{b,t}, s^w_{b, t}, s_{p, t}, s_{b,t})$, where $s_{p, t}$ is $\phi$ in case of an off-platform buyer.

\paragraph{Surplus from a transaction.}
The buyer surplus is $r_{b, t} = \max$ $\{u^w_{b, t}, u^p_{b, t}\}$.
We define the \emph{world matching surplus} and the \emph{platform matching surplus} respectively as 
$$r^w_{b, t}=u^w_{b, t}\cdot \I^w_{b,t} \ \ \text{and} \ \ r^p_{b, t}=u^p_{b, t}\cdot (1-\I^w_{b,t}),$$ 
where $\I^w_{b,t}$ is an indicator of whether buyer $b$ transacted in the world or not at time $t$. 
A seller when chosen by a buyer cannot decline a transaction, and the surplus (i.e., net profit) of seller $s$ is
\begin{equation}
	r_{s, t} = 
	\begin{cases}
		v^1_s(1 - \omega_s - P_{R, k}) & \text{for a  platform transaction,} \\
		v^1_s(1 - \omega_s) & \text{for a  world transaction,}\\
		0 & \text{otherwise.}
	\end{cases}
\end{equation}
We denote $n^p_{s, k}$ the number of transactions completed by seller $s$ via the platform during epoch $k$,  and  $n^w_{s, k}$   the number of transactions completed by the seller in the world.

\subsection{Epoch Surplus and Platform Revenue}
\label{sec:revenue_surplus}
Buyer $b$'s {\em epoch surplus}, $r_{b,k}$, is their total surplus from matching minus any subscription fee paid,
\begin{equation}
    r_{b, k} = \sum_{t \in k} r_{b, t} - \I^p_{b,k} P_{\B, k} = \,\,\underbrace{\!\! \sum_{t \in k} r^w_{b, t}\!\!}_{r^w_{b, k}}\,\, + \,\,\underbrace{\!\! \sum_{t \in k} r^p_{b, t} - \I^p_{b,k} P_{\B, k}\!\!}_{r^p_{b, k}}\,\,,
\end{equation}
where $\I^p_{b,k} \in \{0, 1\}$ indicates whether the the buyer is on or off the platform during epoch $k$. 
This surplus decomposes into surplus from world transactions $r^w_{b, k}$ and surplus from platform transactions $r^p_{b, k}$,
and we define the {\em total buyer surplus generated by the platform} in epoch $k$ as $r^p_{\B, k} = \sum_{b \in \B} r^p_{b, k}$, and the {\em total buyer surplus from the world} in epoch $k$ as $r^w_{\B, k} = \sum_{b \in \B} r^w_{b, k}$.

Similarly, a seller $s$'s {\em epoch surplus}, $r_{s,k}$, is their total net profit from transactions minus any fee,
\begin{equation}
	r_{s, k} = \sum_{t \in k} r_{s, t} - \I^p_{s,k} P_{\S, k} = \,\,\underbrace{\!\! n^w_{s, k} v^1_s(1 - \omega_s)\!\!}_{r^w_{s, k}}\,\, + \,\,\underbrace{\!\! n^p_{s, k} v^1_s (1 - \omega_s - P_{R, k}) -  \I^p_{s,k} P_{\S, k}\!\!}_{r^p_{s, k}}\,\,,
\end{equation}
where $\I^p_{s,k} \in \{0, 1\}$ is an indicator to denote whether the the seller is on or off the platform during epoch $k$. 
We define the {\em total seller surplus from platform transactions} in epoch $k$ as $r^p_{\S, k} = \sum_{s \in \S} r^p_{s, k}$ and the {\em total seller surplus from world transactions} in epoch $k$ as $r^w_{\S, k} = \sum_{s \in \S} r^w_{s, k}$.

The {\em total platform revenue} in epoch $k$ is the sum of the subscription and referral fees it charges,
\begin{equation}
	\label{eq:platform_epoch_profit}
	r_{p, k} = \sum_{b \in \B} \I^p_{b,k} P_{\B, k} + \sum_{s \in \S} \Parens{\I^p_{s,k} P_{\S, k} + n^p_{s, k} v^1_s P_{R, k}}.
\end{equation}

The {\em total welfare} of the economy in epoch $k$ is the sum of all buyer and seller surplus and platform revenue, i.e., $\sum_{b \in \B} r_{b, k} + \sum_{s \in \S}r_{s, k} + r_{p, k}$.

\subsection{Subscription-Level Decisions}
\label{sec:choice_subscribe}
At the start of each epoch, each buyer and seller ``wakes up'' with some probability to reevaluate their current state and decide whether or not to subscribe to the platform.
This decision is utility-theoretic and depends on 
(1) estimating the effect of a subscription decision (i.e., the surplus from joining vs.~operating off platform) 
and (2) an agent-specific inertia that captures the extent to which the agent becomes affiliated with a particular transaction channel. 
We provide in the sequel a high-level description as to how buyers and sellers make such decisions, and defer detailed mathematical expressions to Appendix~\ref{app:counterfactual_estimates}.

\paragraph{Estimating the effect of a subscription decision.}
Each agent, whether a buyer or seller, subscribes by comparing the estimated surplus on and off platform, under the newly proposed platform fees and the observable world transaction friction.
We assume that agents do not coordinate among themselves and thus conduct such comparisons assuming a unilateral change, i.e., being the only one who wakes up and makes a different subscription decision, and that queries from the buyer agent itself as well as others remain the same as in the previous epoch.
For buyers or sellers who were on platform in the past epoch, this estimate means to re-evaluate their transaction decisions under new fees and friction.
For agents who were off platform in the past epoch, we assume that the platform can provide information (honestly) to facilitate this estimation.
In practice such estimates could occur through a trial period offered by the platform or through the platform providing an estimate of costs and benefits based on recent history.

\paragraph{Agent-specific decision inertia.}
We incorporate {\em behavior inertia} in agent choice model to capture an agent's tendency to stick with their current state or decision. 
This kind of inertia has been empirically observed in platform adoption decisions post-pandemic~\citep{oblander2022}, along with many other settings, including choosing consumer packaged goods~\citep{shum2004does,dube2010state} and selecting health and automobile insurance~\citep{handel2013adverse,honka2014quantifying}.

Following prior models \citep{dube2009switching,MacKay2021,farrell1988dynamic}, we include inertia, denoted $\chi$, as an additive term to an agent's surplus from the current decision. 
Specifically, we define inertia as a term with its value growing logarithmically in the number of consecutive epochs for which an agent has made the same choice (either on or off the platform), and ``reset'' this inertia upon a change to the subscription decision. 
Intuitively, this reflects the fact that the longer an agent has stick to a certain option, the less likely they will switch to alternative options.
Based on this adjusted surplus, both buyer and seller agents decide whether or not to pay the fees and join the platform according to probabilities calculated from the standard {\em discrete-choice logit model}~\citep{dube2009switching,MacKay2021} (see  Appendix~\ref{app:counterfactual_estimates} Equation~\eqref{eq:discrete-choice-logit}).

\section{The Platform's Decision Model}
\label{sec:pdp}
%

In this section, we formulate the platform's problem as a POMDP~\citep{KAELBLING1998}, with buyers' knowledge about sellers and transactions completed in the world as private information and thus not observable to the platform. 
The platform learns a fee-setting policy and a matching policy based on observations of the activities of on-platform buyers and sellers.
In our current work, for reasons of computational tractability and interpretability, we choose to study platforms that either follow a typical, myopic optimal matching policy and learn to set fees or follow fixed, regulated fees and learn how to match. 
That said, one of our regulatory interventions is to fix platform fees to a certain regime (e.g., those adopted by a platform in an environment without shocks), studying how the platform subsequently chooses how to match.

\subsection{Learning to Set Fees under Myopic Query Matching}
\label{sec:pricing_policy}
We study the sequential decision making problem of a platform who learns to set fees while using {\em myopic query matching} (i.e., recommending the on-platform seller that yields the highest matching utility to the buyer). 
A POMDP~\citep{KAELBLING1998} can be formally described as a 6-tuple $(\X, \A, \P, \R, \Omega, \O)$, with the state space $\X$, action space $\A$, a Markovian state-action-state transition probability function $\P$, and a reward function $\R$.
Instead of seeing the true state $x \in \X$, the agent receives an observation $o \in \Omega$, generated from the underlying state according to the probability distribution $o \sim \O(x)$.
We describe each component of the POMDP and define the platform's fee-setting policy. 
Here, the platform is making a decision in regard to fees for the upcoming epoch $k$ based on its experience from epoch $k-1$:
%
\begin{itemize}[leftmargin=*]
	\item The {\em state} $x_k \in \X$ at the start of epoch $k$ (i.e., before the platform sets fees) is comprised of
	\begin{enumerate}[label=(\roman*),leftmargin=*]
		\item buyer attributes: the latent location, epoch budget, query distribution, and knowledge of world sellers,
		\item seller attributes: the latent location, cost fraction,  and shutdown threshold,
		\item agent subscription states:  either on- or off-platform for the past epoch, $\I^p_{\B, k-1}$ and $\I^p_{\S, k-1}$,
		\item the agent inertia levels: $\chi_{b,k-1}$ and $\chi_{s,k-1}$,
		\item a sequence of query, seller candidates, and buyer's choices in previous epoch: $Q_{k-1} = \{(q_{b,t}, s^p_{b,t}, s^w_{b,t}, s_{b,t})\}_{t \in k-1, \ b \in \B}$,
		\item the shutdown states for sellers: whether a seller has shut down at the end of epoch $k-1$, $\I_{\S, k-1}$,
		\item the platform fees for the past epoch: $P_{\B, k-1}, P_{\S, k-1}, P_{R, k-1}$,
		\item the world transaction friction for the current epoch: $\mu_k$.
	\end{enumerate}
	\item An {\em action} $a_k = (P_{\B, k}, P_{\S, k}, P_{R, k})$ defines the  fees for the upcoming epoch $k$. We model a discrete action space $\A$ where fees take discrete values at integer multiples of a tick (or percentage) size.
	\item For the {\em state transition} $\P: \X \times \A \rightarrow \Delta(\X)$, agent attributes (e.g., latent locations, knowledge of sellers) remain the same across epochs. 
 Buyers and (viable) sellers follow their choice model to subscribe to the platform (Section~\ref{sec:choice_subscribe}), leading to new subscription states and inertia levels. 
	For each time step $t \in k$, we follow the ``query, match, transact'' dynamics (Section~\ref{sec:dynamics}), which gives a full sequence of $Q_k$.
	Each viable seller may shut down based on the surplus in epoch $k$ and their shutdown threshold.
	Fees follow from the actions taken, and the world transaction friction evolves according to a defined Markov process.
	Altogether, this gives a new state $x_{k+1} \sim \P(x_k, a_k)$.
	\item A {\em reward} $r_k \sim \R(x_k, a_k)$ is provided to the platform at the end of epoch $k$, when agent subscription and transaction outcomes are available. 
	The reward can be set to model different platform objectives, integrating considerations that come from regulation. 
	%
	\item The platform's {\em observation} $o_{k} \in \Omega$ consists
	of the world transaction friction $\mu_k$, agent subscription states in the past epoch, and the sequence of queries generated by on-platform buyers, as well as their decisions on whether or not to transact via the platform, i.e., 
	$Q^p_{k-1} := \{(q_{b,t}, s^p_{b,t}, \ind \{s_{b,t} = s^p_{b,t}\})\}_{t \in {k-1}, \ b \in \B_{k-1}}$
 (and not counterparties in off-platform transactions). 
\end{itemize}
\subsection{Learning to Match Queries under Fixed, Regulated Fees}
\label{sec:matching_policy}
In a second setting, we study the sequential decision making problem of
a platform who learns how to match buyer queries with on-platform sellers when the platform fees are fixed, for example
due to regulation.
Typical myopic matching, as described above, favors the buyer side of the market, by directing a query to the buyer's utility-maximizing, on-platform seller.
To complement this, we model  matching strategies that can choose to benefit other parties in the economy, for example sellers or the platform itself.
To facilitate interpretability, we define a {\em matching strategy} by two parameters: (1)~a {\em matching utility threshold} $\eta \in [0, 1]$, that specifies the minimum utility that a recommended seller should provide to the buyer, as a fraction of utility from the myopically-optimal match, 
and (2) a {\em matching rule}, which  directs how to pick a seller amongst those that meet this utility threshold.
We  consider two rules: 
\begin{itemize}[leftmargin=*]
	\item {\em The seller-aware matching rule}: Among sellers who meet the utility threshold,
	match a query to the seller who has the lowest surplus on the platform so far during the epoch,%
	\footnote{We assume  for simplicity that the platform knows a seller's production cost, and thus can calculate its surplus from transactions. In practice, this idealized seller-aware rule could be modified and defined in terms of the number of platform transactions completed by a seller, or other  inferred quantities about seller surplus.} 
	breaking ties in the buyer's favor.
	\item {\em The profit-driven matching rule}: Among sellers who meet the utility threshold,
	match a query to the seller who brings the largest revenue to the platform, breaking ties in the buyer's favor  (given referral fees in our setting, this is the most expensive, on-platform seller). 
\end{itemize}
%
%
As a special case,  myopic query matching corresponds to setting the matching threshold $\eta =1$ (along with either rule, since they have no choice as to how to further select a seller).

Intuitively, the seller-aware rule may be useful in promoting a more diverse set of sellers, by increasing sales to those who have been benefiting less from the platform,  
whereas the profit-driven rule is at the other end of the spectrum, aiming to maximize the platform's myopic transaction revenue without considering longer-term effects.
We detail the implementation of a matching strategy, given the matching rule and a utility threshold in Appendix~\ref{app:pdp} \Algo{greedy_matching}.

The goal of the platform is to learn a {\em matching policy} that chooses a matching strategy for \textit{each epoch}---a utility threshold and a matching rule, based on an observation as to which buyers and sellers choose to subscribe for the \textit{upcoming} epoch.
To model this, we make several adjustments to the fee-setting POMDP in defining a {\em matching POMDP}, as in this case, the platform observation needs to be defined \textit{after} agents make subscription decisions.
We defer detailed adjustments to Appendix~\ref{app:matching_policy_app}.\looseness=-1
%

\subsection{Finding the Optimal Policy under Regulatory Interventions}
Interactions between the platform and buyers and sellers can be considered as a {\em Stackelberg game}: the platform agent is the {\em leader}, choosing the fee or matching policies, and buyers and sellers are the {\em followers}, responding to platform policies. 
In our environment, the buyer and seller strategies are  a fixed mapping from prior matching experience, platform fees, and  world transaction friction to decisions in regard to joining or exiting the platform,  and we can handle this Stackelberg structure by modeling the agents within the POMDP of the platform (i.e., as part of the transition model).
%

Depending on the setting, the platform learns a fee policy or a matching policy, denoted $\pi(a | o_k)$, to maximize its discounted cumulative reward across different episodes.
An optimal policy in a POMDP requires the action to be taken depending on the entire history of observations, which we denote $h_k := \{o_0, a_0, ..., a_{k-1}, o_k\}$.
Following the success of using representation learning to solve POMDPs~\citep{Heess2015,Wierstra2007,Hausknecht2015}, we use neural networks to learn sufficient statistics of the history, denoted $T\!(h_k)$.
Altogether, we use deep RL to learn the platform policy $\pi(a | o_k; \vtheta)$, specifically parameters $\vtheta$ of a neural network that extract $T\!(h_k)$ and map to actions to maximize the platform objective,
\begin{equation}
	\label{eq:rl_objective}
	\max_{\pi} \quad \E_{a \sim \pi, x \sim \P} \left[\sum_{k=0}^{K} \gamma^k r_k \right],
\end{equation}
where $\gamma\in (0,1)$ is the discount factor, $K = |\tau|$ is the total number of epochs in an episode, and $r_k$ is based on Equation~\eqref{eq:platform_epoch_profit} and whose precise value can further depend on a regulatory structure, for example a taxation policy. 
We discuss the algorithm that we use to solve this POMDP and give additional implementation details in Section~\ref{sec:experiments}.

\section{Platform Behavior under Shocks and Regulations}
\label{sec:experiments}

We study platform behavior and its effect on the economic system under the following regulatory interventions: (i)~taxation policies that directly change the reward to the platform, (ii)~fee caps that are enacted through restricting the platform's action space, and (iii)~fee freezes that are studied along with a platform who can continue to change its matching policy.
We first provide specifics on the configuration of the simulation settings for our experiments and present main experimental results.\looseness=-1

\subsection{Environment Settings of the Platform-Based Economy}
\begin{figure}[t]
	\centering
	\begin{subfigure}{0.32\columnwidth}	
		\centering
		\includegraphics[width=0.99\columnwidth]{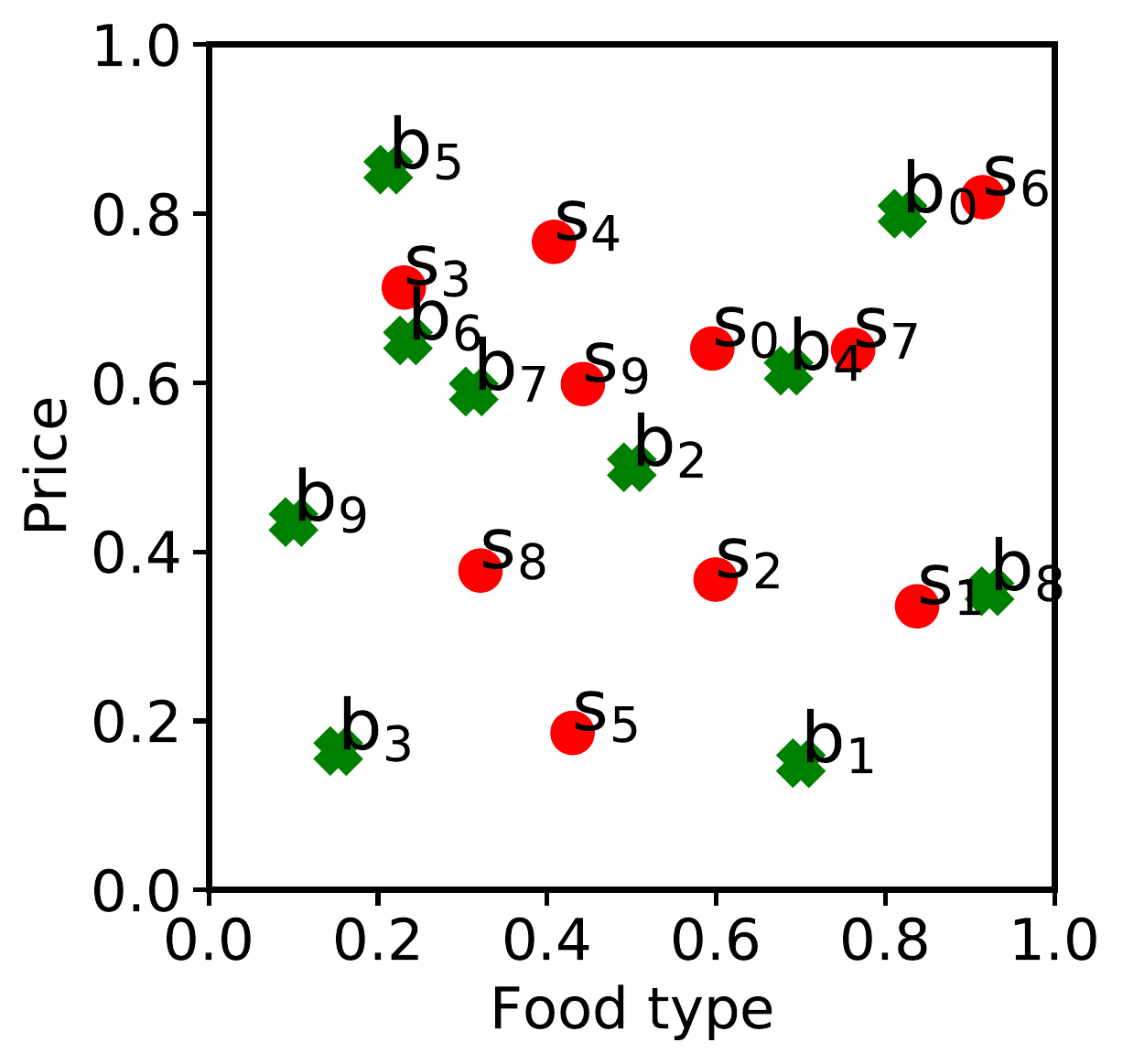}
		\caption{Uniform.}
	\end{subfigure}
	\hspace{0.005\columnwidth}
	\begin{subfigure}{0.32\columnwidth}	
		\centering
		\includegraphics[width=0.99\columnwidth]{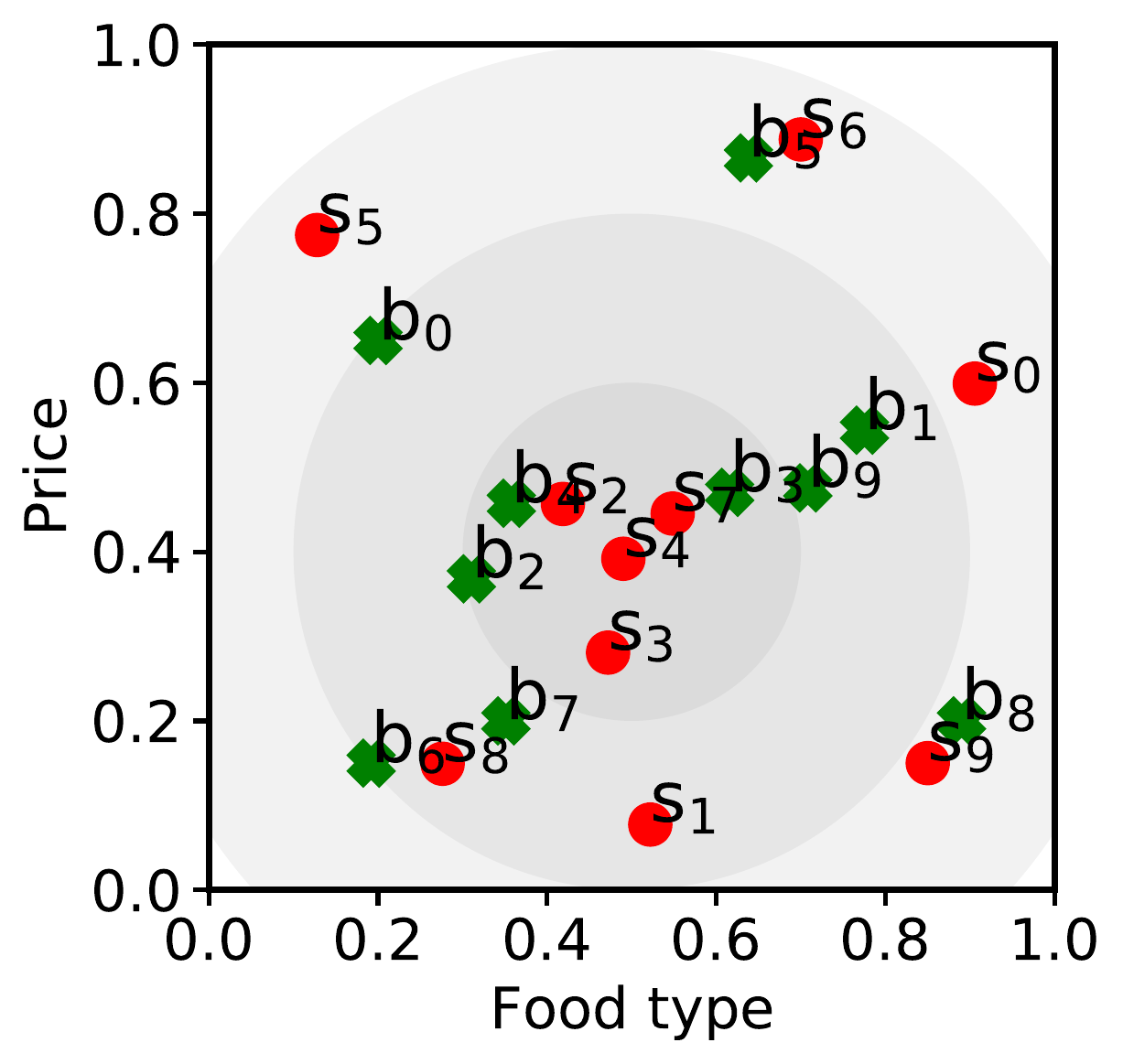}
		\caption{Core-and-Niche.}
	\end{subfigure}
	\hspace{0.005\columnwidth}
	\begin{subfigure}{0.32\columnwidth}	
		\centering
		\includegraphics[width=0.99\columnwidth]{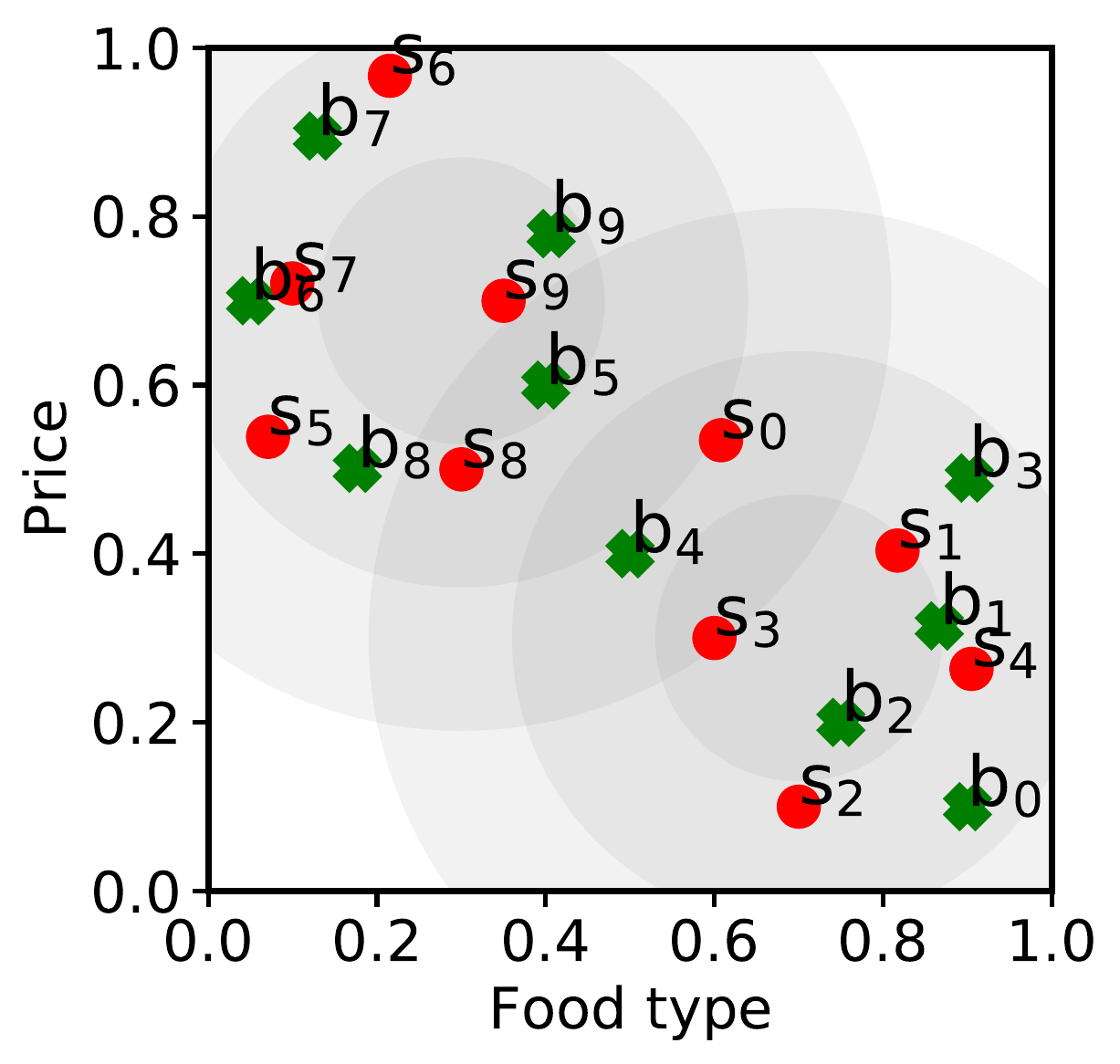}
		\caption{Two-Core.}
	\end{subfigure}
	\caption{An illustration of latent locations of buyers and sellers for each market structure.}
 \vspace{-2ex}
	\label{fig:three_structures}
\end{figure}

\paragraph{Market structure and dynamics.}
We follow Section~\ref{sec:gym_env} in specifying a range of different market environments.
We consider three types of \textit{market structures}, corresponding to distinct latent locations of buyers and sellers: 
\begin{itemize} [leftmargin=*]
	\item \emph{Uniform:} Buyers and sellers are uniformly distributed, i.e., $v \sim U[0, 1]^2$, representing an economy with diverse buyer interests and seller attributes. 
	\item \emph{Core-and-Niche:} There is a propensity for buyers and sellers to locate towards the center, i.e., $v \sim \text{truncated Gaussian}(\mu, \sigma, 0, 1)$ with $\mu = [0.5, 0.4]$ and $\sigma = 0.2$, representing a ``core'' of agents around $\mu$ as well as ``niche'' agents located away from the market center. 
    We choose the distribution to be slightly biased towards lower prices, representing the majority.
	\item \emph{Two-Core:} 
	One group $v\sim \text{truncated Gaussian}(\mu_1, \sigma_1, 0, 1)$ with $\mu_1 = [0.7, 0.3]$ and $\sigma_1 = 0.17$, and another group $v\sim \text{truncated}$ $\text{Gaussian}(\mu_2, \sigma_2, 0, 1)$ with $\mu_2 = [0.3, 0.7]$ and $\sigma_2 = 0.17$. 
The first  group  of buyers and sellers is centered around relatively cheap options, and the second around more expensive ones.
\end{itemize}
Across all market structures, we consider environments with 10 buyers and 10 sellers, an episode that consists of $K=12$ epochs, with each epoch containing $T=100$ time steps. 
Figure~\ref{fig:three_structures} provides a visualization for each market structure.
For each latent location, we also consider markets that vary in regard to the buyer's knowledge level about sellers, denoted $\rho$. 
In particular, each buyer $b$ samples i.i.d. 
 $\mathit{Bern}(\rho)$ for each seller, to generate its set of known sellers $\S_b$.

\paragraph{Agent attributes.}
Buyers arrive in a round robin fashion.
A buyer $b$ who arrives at $t$  submits a query around the buyer's 
 latent location, $q_{b,t} \sim \N(v_b, \sigma^2_b)$, with $\sigma^2_b = 0.02$. 
When query $q$ is fulfilled by seller $s$, the buyer receives a matching utility of $u_\B(q, s) = \exp(-c\norm{q - v_s}_2)$, where we choose $c=2$ to have matching utilities span $[0, 1]$. 
%
We set a buyer's \emph{per-epoch budget} $\psi_b$ to be $v_b^1$ (their price preferences) times the number of arrivals within an epoch.
%
Each seller's fractional cost is randomly drawn from $\omega_s \sim U[0.2, 0.4]$.
We set the shutdown threshold to $\lambda=2$ for all sellers, meaning that a seller will go bankrupt after two consecutive epochs without any (positive) surplus.
Each agent, whether a buyer or seller, is initialized randomly to stay in the world or join the platform.

\begin{wrapfigure}[12]{r}{0.45\textwidth}
	\centering
    \vspace{-2ex}
	\includegraphics[width=0.35\columnwidth]{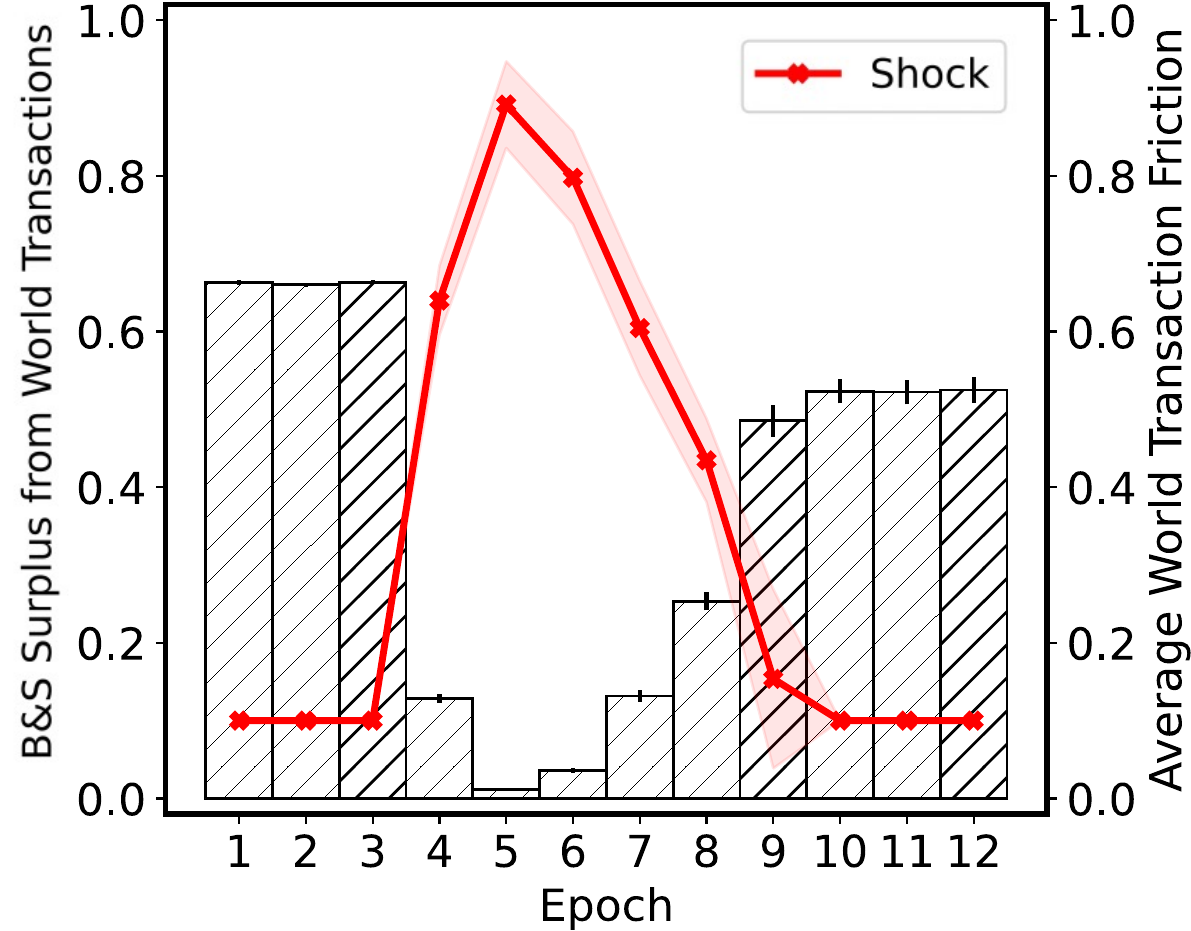}
 \vspace{-1ex}
	\caption[width=0.4\columnwidth]{\small Normalized agent surplus (bars) in markets with no platform under a full cycle of shock.\looseness=-1
    \label{fig:welfare_decomposition}}
\end{wrapfigure}

\paragraph{Shocks.}
We vary the world transaction friction, $\mu_k$, across epochs to model the {\em pre-, during}, and {\em post-shock} stages.
To facilitate the pre- and post-shock comparison, we set the pre- and post-shock stages to each last for at least three epochs and with low world friction, $\mu_k = 0.1$.
The shock stage is controlled by a {\em shock intensity}, $I \sim U[I_{\min}, I_{\max}]$, specifying the largest value that can be attained.  
We sample $\mu_k \sim$ Lognormal$(\mu=0, \sigma=0.5)$, and multiply the values by the intensity $I$. 
%
Figure~\ref{fig:welfare_decomposition} (red line) shows the average shock schedule across a hundred simulated episodes with $I \sim U[0.8, 1]$.

\paragraph{Platform action space.}
For a platform agent that sets fees, its action space is divided into three subspaces, one for each fee type.
Without caps or regulations on fees, the registration fees, $P_{\B, k}$ and $P_{\S, k}$,  range from 0 to 10, with discrete levels at intervals of 0.2. 
The seller referral rate, $P_{R, k}$, ranges from 0 to 1, with discrete levels at intervals of 0.1.

For a platform agent that decides how to match queries to sellers,
the action space is a combination of the choice of matching rule (seller-aware or profit-driven) and a matching utility threshold, ranging from 0 to 1 and discretized at intervals of 0.1. 

\paragraph{Implementation details.}
We sample initial states with a {\em warm-up epoch} (in addition to the twelve epochs), where the friction is set to $\mu_0=0.1$, the platform charges no fees, and buyers and sellers join the platform based on their initialization. 
The experience gained by buyers and sellers in this warm-up epoch provides a basis for the platform to choose actions, and for the buyers and sellers to form estimates to guide their subscription decisions. 

\smallskip
Following the platform POMDP formulation (Section~\ref{sec:pdp}), we make the following observations available to the platform:
\begin{enumerate}[label=(\roman*),leftmargin=*,itemsep=0ex]
	\item On-platform buyers and sellers, represented by two binary vectors indicating their subscription states, i.e., $\I^p_\B$ and $\I^p_\S$,
	\item Summary statistics of on-platform agents, including the number of platform transactions and the surplus accumulated on platform so far in the epoch,
	\item The platform matching and transaction matrix between {\em on-platform} buyers and on-platform sellers for the epoch,
	\item The platform fees, the matching rule and utility threshold (if the platform learns matching), and the current epoch's world friction.
\end{enumerate}

\begin{figure}[t]
	\centering
	\includegraphics[width=0.8\columnwidth]{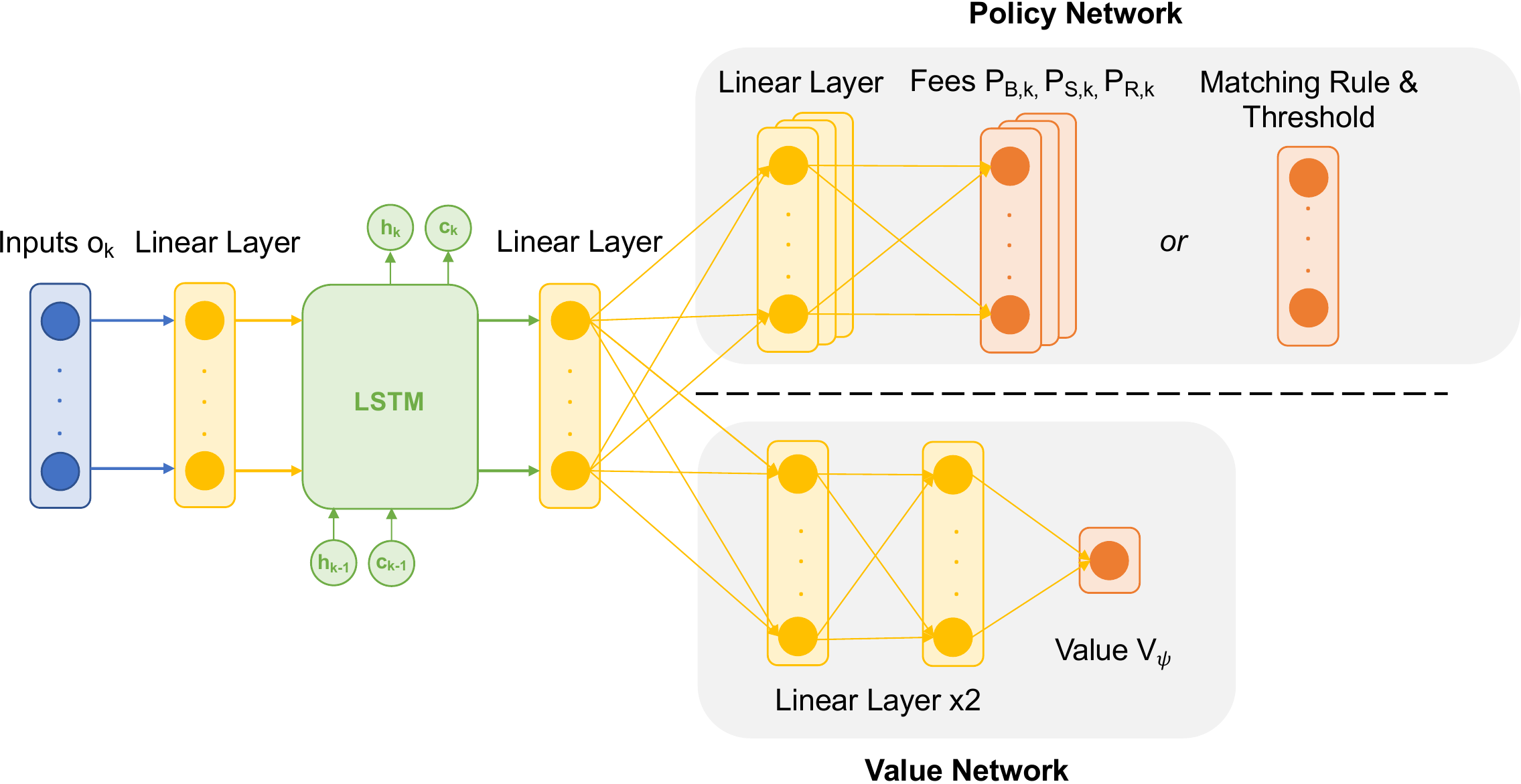}
	\caption{Neural network structure for the platform fee-setting and matching policies.
	\label{fig:NN_diagram}}
\end{figure}

We use the {\em Advantage Actor-Critic (A2C)} algorithm~\citep{rlbook,Degris2012} to learn the optimal platform policy.
%
%
Based on preliminary explorations, we design the actor and critic to share a fully-connected layer, LSTM cells of size 128, and again a fully-connected layer to recover sufficient statistics of the history, using this to in effect infer the knowledge structure of buyers about sellers and the demand elasticity of agents to platform fees under shocks.
Each network also has its own two fully-connected layers. 
The critic outputs the value $V_\psi(o)$ of an observation $o$, and the actor gives policy $\pi_\theta$ for an observation $o$. 
For the fee-setting actor, this includes three separate output layers, with each returning a vector of probabilities for one type of platform fee.
For the matching actor, this is a vector of probabilities over the matching utility thresholds that are applicable to both matching rules. 
Figure~\ref{fig:NN_diagram} illustrates the neural network structure we implement.

Besides policy gradient loss, we apply {\em entropy regularization} to the policy network to encourage exploration~\citep{Mnih2016,Williams1991}.
The respective losses for the policy network and the value network are,
$$\mathcal{L}_{\pi} = -\log \pi(a_k | o_k; \vtheta)(R_k - V_\psi(o_k)) - \beta \mathcal{H}(\pi(A_k | o_k ; \vtheta)) \ \text{ and } \ \mathcal{L}_{V} = (R_k - V_\psi(o_k))^2,$$
where $R_k$ is the observed return and $\mathcal{H}$ denotes the entropy over learned action probabilities.

%

We tune the platform policy with various combinations of learning rates $\{0.0001, 0.0005, 0.001\}$, batch sizes $\{4, 16, 32, 64, 128\}$, and entropy weights $\{0.001, 0.01, 0.05\}$, and select hyperparameters that maximize the objective function. 
In experiments, we present the average performance across a hundred test episodes from two models trained with different torch seeds.
We report the detailed training parameters in Appendix~\ref{app:experiments}.

\begin{figure}[t]
	\centering
	\begin{subfigure}{0.46\columnwidth}	
		\centering
		\includegraphics[width=\columnwidth]{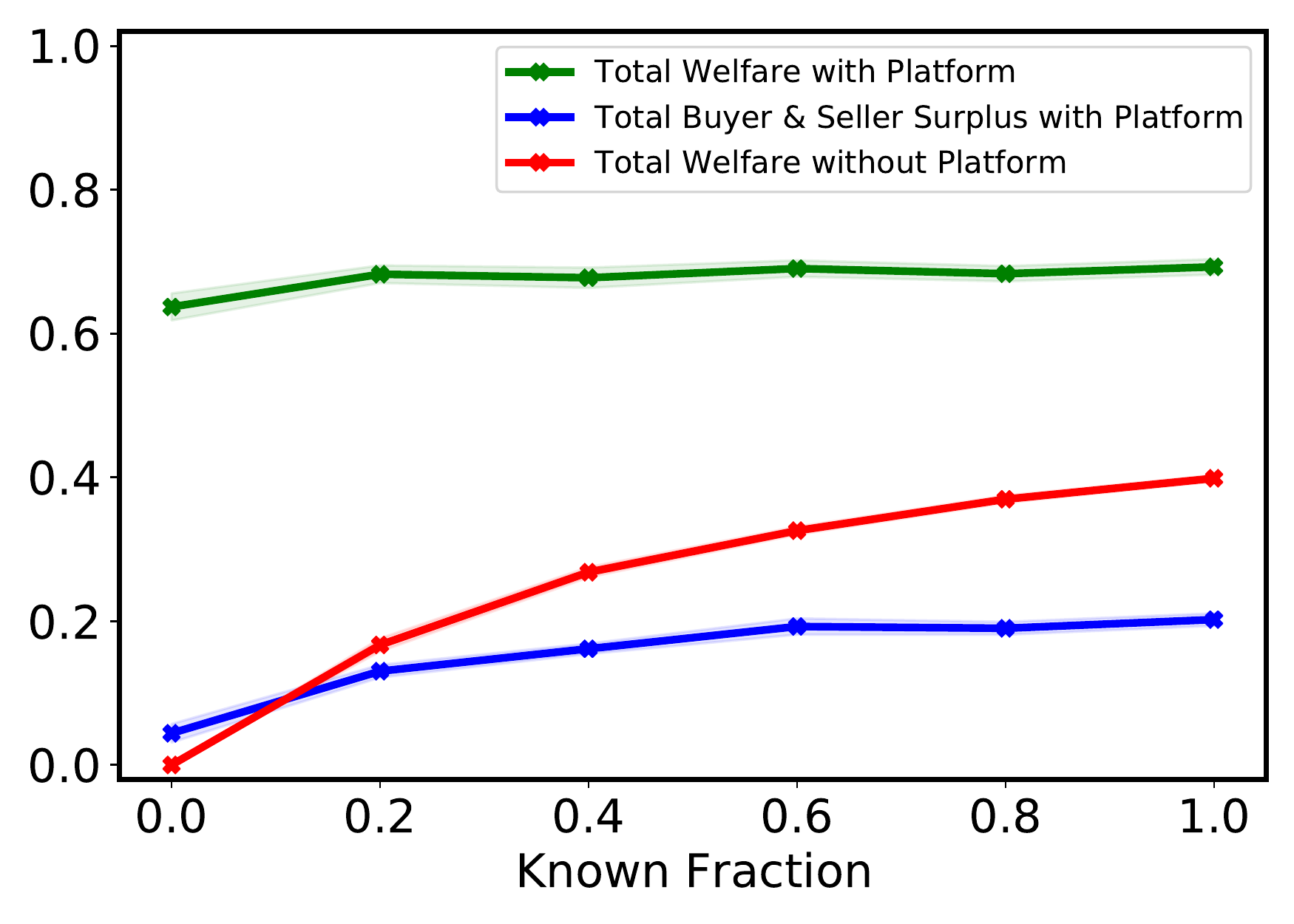}
		\caption{Varying $\rho$ buyers' knowledge level about sellers, with fixed $\mu = 0.6$.}
	\end{subfigure}
	\hspace{0.03\columnwidth}
	\begin{subfigure}{0.46\columnwidth}	
		\centering
		\includegraphics[width=\columnwidth]{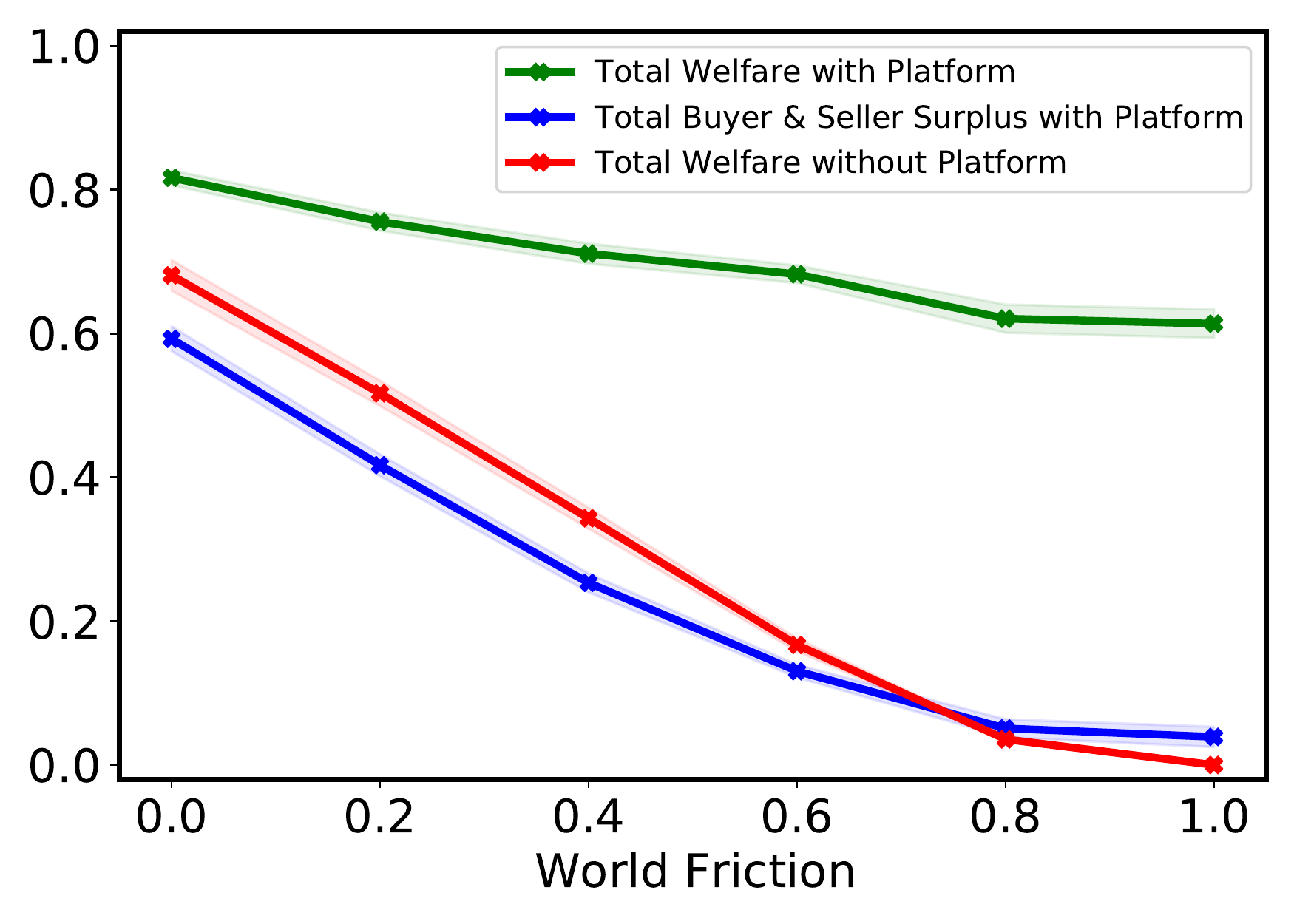}
		\caption{Varying $\mu$ the world transaction friction, with fixed $\rho = 0.2$.}
	\end{subfigure}

	\caption{Total welfare and buyer and seller surplus achieved in environments that vary in $\rho$ and $\mu$, with and without a platform under the Core-and-Niche market structure. 
		\label{fig:platform_value_}}
\end{figure}

\subsection{Characterizing the Environment: Value Created by the Platform}
We start by building some intuition about the basic economics of our simulation environments.
For this, we characterize the value generated by a revenue-maximizing platform across a range of \emph{single-epoch, no-shock} environments that respectively vary in (1)~buyers' knowledge level about sellers $\rho$ and (2)~the world transaction friction $\mu$. 
The warm-up epoch is retained to facilitate agents' subscription decisions.
For each market structure, we generate three samples of latent locations of buyer and seller agents, and for each latent sample and knowledge level, $\rho$,  we sample ten knowledge structures, specifying which sellers are known by each buyer. 
For these single-epoch baseline settings,
we can use {\em Bayesian Optimization} (BO)~\citep{bo} in place of RL to find platform fees that maximize the platform's revenue, and conduct controlled experiments with and without a platform. 

Figure~\ref{fig:platform_value_} shows the buyer and seller surplus, as well as the total welfare of the Core-and-Niche markets.%
\footnote{We defer results for the other two market structures to Appendix~\ref{sec:apx-other_latents}, which are qualitatively similar to the presented Core-and-Niche market.} 
In the case where there is no platform, the total welfare is equal to the total buyer and seller surplus.
We normalize surplus by the total \textit{ideal welfare} achieved in a world where each buyer knows all sellers and there is no world transaction friction (or this is equivalent to a benevolent platform that charges no fee).
We validate the simulator by confirming that under no platform scenarios (Figure~\ref{fig:platform_value_}, red lines), total welfare increases as buyers' knowledge level about sellers increases and the world friction decreases.

Across all environments varying in $\rho$ and $\mu$, a revenue-maximizing platform consistently increases total welfare relative to the absence of a platform, creating value by reducing search costs (i.e., matching buyers to unknown sellers) and facilitating transactions (i.e., reducing off-platform fulfilment costs). 
We can also confirm that the revenue a platform can extract (i.e., differences between green and blue lines) increases as buyers have less knowledge about sellers and as the world transaction friction increases.
For scenarios when $\rho$ is extremely low or $\mu$ is very high, the platform possesses large market power and can extract all surplus from buyers and sellers.

\subsection{Platform Responses to Interventions under Market Shocks}

We now present our main results, analyzing in the presence of market shocks, a no-intervention case as well as three different kinds of interventions. 
The headline result is that, whereas taxes, and caps imposed on some subset of fees, are largely ineffective,  there are two forms of interventions---(1)~capping all fees while leaving the platform the flexibility to continue to set fees subject to these caps and (2)~fixing fees to those that a self-interested platform chooses in an environment \textit{without} shocks while allowing the platform to continue to optimize its matching policy---that can successfully incentivize a revenue-maximizing platform to promote the seller diversity, efficiency, and resilience of the overall economy.
We find the second kind of successful intervention especially interesting because it seems of particular relevance to practice (the knowledge of how to set fees comes from the platform's own behavior, and the platform can continue to flexibly make matching decisions). 
More broadly, we aim to bring the strategic aspects of platform matching to attention in considering different regulatory interventions and evaluating their effectiveness in incentive alignment.
\paragraph{Case 1: Platform fee setting in a laissez-fair system.}
\begin{figure}[t]
    \centering
    \includegraphics[width=0.8\columnwidth]{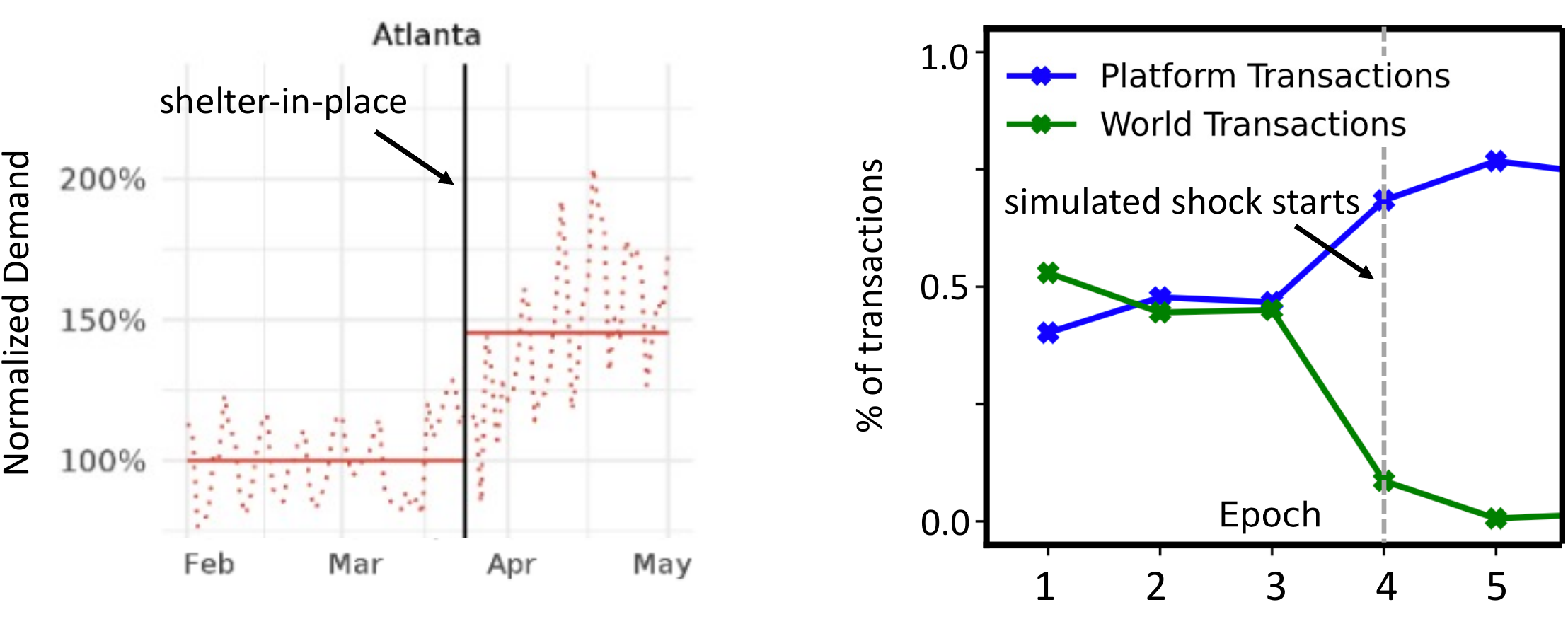}
    \caption{A comparison of empirically-observed demand surge after the shelter-in-place order as shown in~\citet{raj2021} (Left) to the increase in the number of on-platform transactions induced by our simulated economic shock (Right).
    \label{fig:my_label}}
\end{figure}
\begin{figure}[b!]
	\centering
	\begin{subfigure}{\columnwidth}	
		\centering
		\includegraphics[width=\columnwidth]{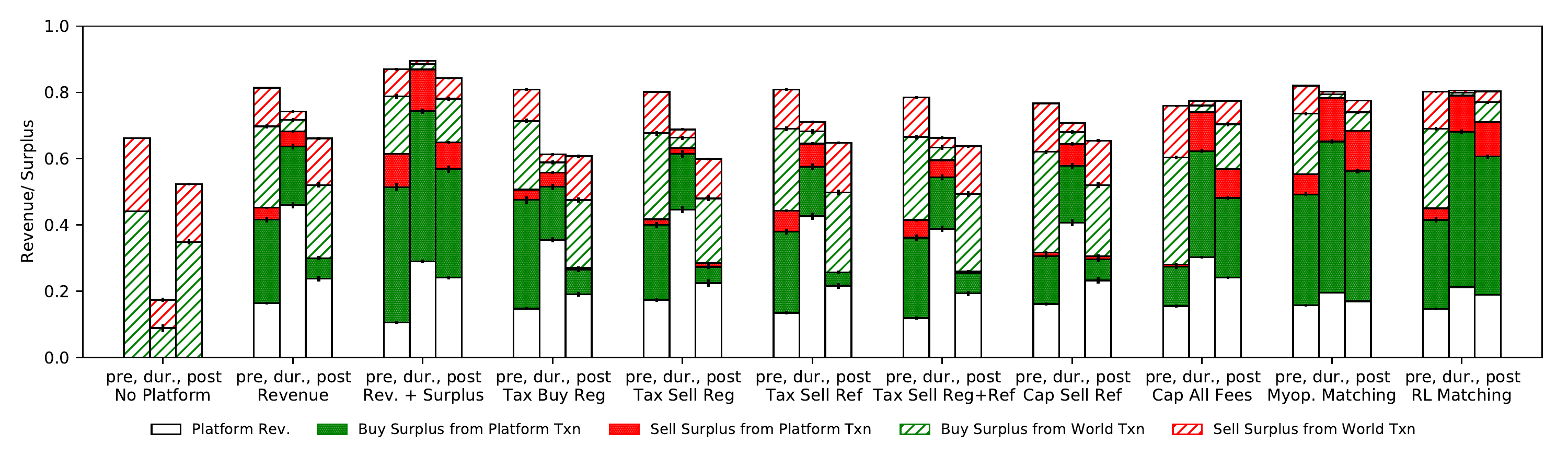}
		\caption{Welfare decomposition achieved by different learned fee-setting policies in each shock stage.}
            \vspace{2ex}
		\label{fig:all_obj_results_profit_surplus}
	\end{subfigure}
	\begin{subfigure}{\columnwidth}	
		\centering
		\includegraphics[width=\columnwidth]{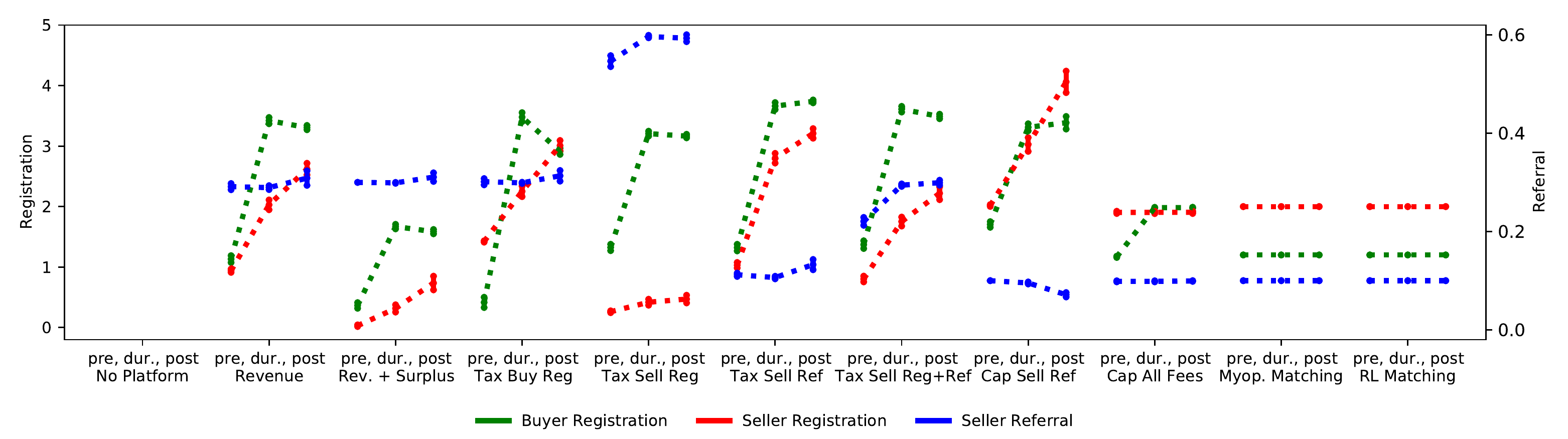}
		\caption{Platform fees set for each shock stage.}
            \vspace{2ex}
		\label{fig:all_obj_results_price}
	\end{subfigure}
        
 	\begin{subfigure}{\columnwidth}	
		\centering
		\includegraphics[width=\columnwidth]{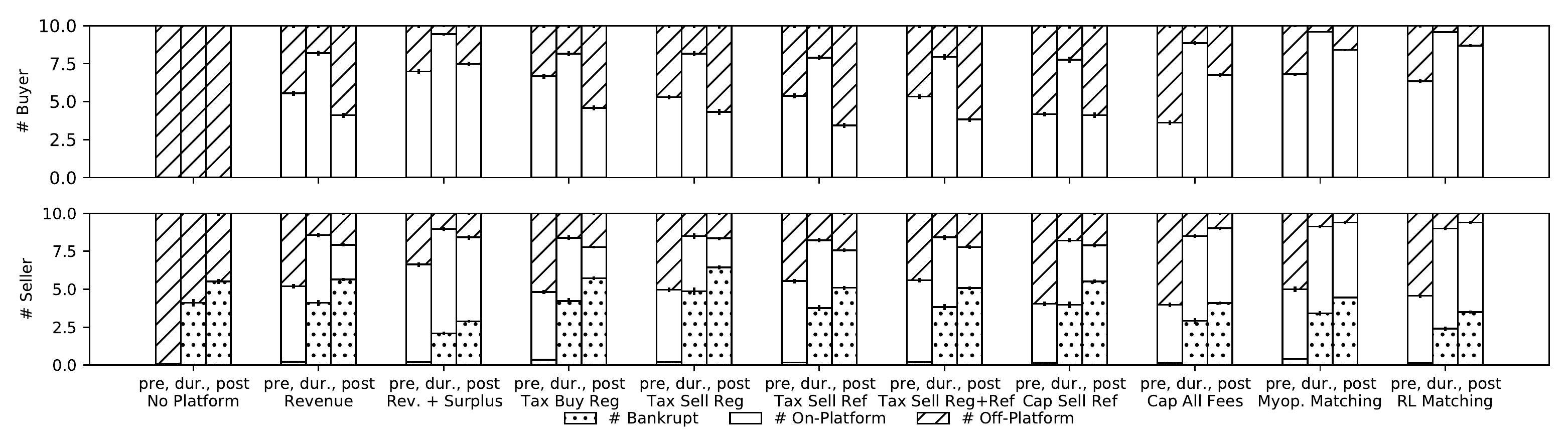}
		\caption{The number of on- or off-platform agents and bankrupt sellers induced by different regulatory interventions.}
		\label{fig:all_obj_results_price}
	\end{subfigure}

	\caption{The welfare decomposition, platform fees, and agent states induced by different design objectives and considerations coming from regulation under the Core-and-Niche market structure.}
	\label{fig:all_obj_results}
\end{figure}

We first consider a platform that is free from any form of economic interventions:
it adopts \textit{myopic query matching} while learning to set fees to maximize its revenue.
In Figure~\ref{fig:my_label}, we first illustrate the similarity between our simulated increase in the number of platform transactions as an economic shock occurs to the empirically-observed demand surge in Atlanta after the COVID-19 shelter-in-place launched in late March, which is representative of changes in other cities (i.e., on-platform transactions, or demand, increasing by around 50\% during the shock). 
This builds a basic validation of the economic dynamics in our model.\looseness=-1
%

Figure~\ref{fig:all_obj_results}, columns 1-2, compares the outcomes achieved in markets without and with a platform.
%
%
%
As a first observation, from Figure~\ref{fig:all_obj_results}a, we see that the platform generally improves the overall economic welfare, considering the sum of the revenue to the platform and the surplus to buyers and sellers (i.e., the total heights of the bars).
This is especially salient during the shock stage (i.e., central bars), when the world transaction friction becomes high and very few transactions can take place and generate surplus in the absence of a platform. 

At the same time, we see two less beneficial outcomes. 
First, by comparing pre- and post-shock periods (i.e., the left and right bars of the same column), we find that in both settings with and without a platform, the overall economic welfare falls after the shock.
This arises as a result of sellers going bankrupt during the shock, and thus buyer queries can only be matched to less-preferred sellers afterwards.
As further verified in Figure~\ref{fig:all_obj_results}c, columns 1-2, we see that the number of bankrupt sellers in markets with a platform is almost the same as that in markets without a platform, indicating the ineffectiveness of adopting a self-interested, revenue-maximizing platform in helping businesses to survive an economic shock.
In addition, by comparing the surplus to buyers and sellers with and without the platform, we find that the platform tends to reduce total surplus to buyers and sellers in the post-shock stage relative to the case without a platform. 
%

%

We further characterize the type of seller that is more likely to go bankrupt  under markets with and without a platform,
 classifying sellers into three groups: {\em core sellers} that are located within one standard deviation of the center and with at least two buyers nearby, 
{\em niche sellers} that are beyond two standard deviations from the center with at most one buyer nearby,
and {\em cheap sellers} with prices in the lower quartile. 
%
%
As shown in Table~\ref{table:seller_stats}, cheap sellers are much more likely to go bankrupt in markets with a platform, whereas niche ones have higher bankrupt probability  in settings without a platform. 
This suggests that a rational platform  learns to raise fees as much as possible during the shock (Figure~\ref{fig:all_obj_results}b), causing sellers who have lower profit margins and cannot afford the fees to quit the market.
As a result, buyers who prefer cheap options are forced to match with expensive sellers.
\paragraph{A surplus-aware platform.}
As an idealized baseline for regulatory interventions, we can consider the effect of a \textit{surplus-aware platform}, that sets fees to optimize some  combination of its own revenue and the on-platform user (buyer and seller) surplus, i.e., $r_{p,k} + \alpha(r^p_{b, k} + r^p_{s, k})$ and here we choose $\alpha=0.4$.
As shown in Figure~\ref{fig:all_obj_results} (column 3), this platform would in effect restore a comparable level of overall economic welfare after the shock, and lead to fewer bankrupt sellers (see Table~\ref{table:seller_stats}).

\begin{figure}[t]
	\centering
	\includegraphics[width=\columnwidth]{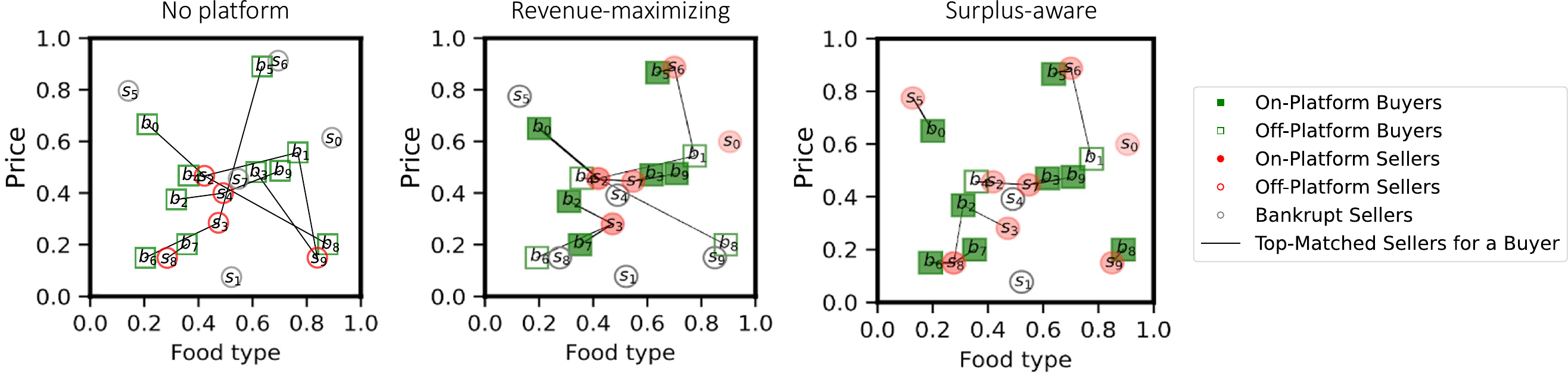}
	\caption{Snapshots of the post-shock market scenarios of economic systems with no platform, a revenue-maximizing platform, and a surplus-aware platform.
	\label{fig:micro_core_bankrupt}}
\end{figure}

In Figure~\ref{fig:micro_core_bankrupt}, we capture snapshots of agent states and top-matched sellers in the last epoch (i.e., post-shock) of markets with no platform, a revenue-maximizing platform, and the surplus-aware platform.
Controlled experiments are conducted for these settings, so that any change in agent behavior is caused by the presence or different fee-setting policies of the platform.
As one can see, for the economic system without a platform, niche sellers go bankrupt, whereas for the revenue-maximizing platform, more cheap sellers go bankrupt, resulting in buyer queries to be matched to less-preferred sellers (e.g., queries from $b_0$ and $b_8$ are matched to $s_2$). 
When the platform also cares about user surplus, fewer sellers, often those with less demand, shut down.
\begin{table}[t]
		\centering
            \small
		\begin{tabular}{lllll}
			\textbf{Type of seller} & \textbf{No platform} & \textbf{Revenue-maximizing} & \textbf{Surplus-aware} \\
			\hline\midrule
			All sellers & 0.55 (0.02) & 0.57 (0.01) & 0.29 (0.01)\\[0.3em]
			Core sellers & 0.33 (0.03) & 0.46 (0.02) & 0.28 (0.02)\\[0.3em]
			Niche sellers & 0.88 (0.08) & 0.45 (0.03) & 0.17 (0.02)\\[0.3em]
			Cheap sellers & 0.52 (0.04) & 0.81 (0.03) & 0.43 (0.03)\\\hline
		\end{tabular}
	\caption{Seller bankrupt frequencies in markets with no platform, a revenue-maximizing platform, and a surplus-aware platform. 
	\label{table:seller_stats}}
\end{table}
%

In the sequel, we consider more practical interventions and evaluate their effectiveness under optimal platform responses.

\paragraph{\textbf{Case 2: platform fee setting under taxation policies.}}
We study the effect of taxation policies imposed on platform profits made from different categories of fees,  with tax schemes that charge a 20\% tax on profits made from buyer registration fees, seller registration fees, seller referral fees, and all fees from sellers respectively.%
\footnote{In experiments, we have tried a range of tax rates (i.e., 20\%, 40\%, 60\%). Despite differences in the absolute fee values, they lead to qualitatively similar trends.} 
As shown in Figure~\ref{fig:all_obj_results}, columns 4-7, our main finding in this regard is that taxation policies in general lead to similar outcomes as those observed in the laissez-fair system (column 2): the overall economic welfare decreases after the shock due to bankrupt sellers.
Specifically, charging taxes on a particular kind of fee leads a rational platform to increase other fees, transferring loss to another user group. 
For example, we compare fees set by a platform under which taxes are imposed on seller referral profits (Figure~\ref{fig:all_obj_results}b, column 6) to platform fees set in a laissez-fair system: the platform learns to decrease seller-referral rates while raising registration fees during and after the shock.
As a result, on-platform sellers achieve higher surplus (see Figure~\ref{fig:all_obj_results}a, column 6, solid red) relative to the laissez-fair case, however, at the expense of lower on-platform buyer surplus (solid green).
In another case, where tax is charged on seller registration revenue (Figure~\ref{fig:all_obj_results}b, column 5), the platform chooses to significantly increase seller referral rates, leading to even lower surplus to platform sellers.


\paragraph{\textbf{Case 3: platform fee setting under fee caps.}}
As a third case, we first study the introduction of a 10\% cap on the referral fee  
(the type of intervention that has been launched in several states in the U.S.).
Similar to the case of taxes imposed on referral profits, the platform learns to respond by raising the registration fees (Figure~\ref{fig:all_obj_results}b, column 8). 
This leads more buyers and sellers to stay off the platform, especially before the shock, and thus reduces the on-platform surplus.

As an example of a more positive kind of intervention, we further consider the case where \textit{every} kind of fee is capped, here with $P_{\B, k} \leq 2$, $P_{\S, k} \leq 2$, and $P_{R, k} \leq 0.1$.
As shown in Figure~\ref{fig:all_obj_results}b, column 9, under these fee caps, the platform chooses to set the maximum possible fees, except for a lower value in the case of the pre-shock buyer registration. 
We find that such caps in effect may induce platform fee-setting policies that benefit the economic system: the overall welfare is not affect by the shock, with more viable sellers after the shock offering good query matches. 
This leads to the intervention that we study in Case 4, which operates without needing special knowledge of how to cap fees and still provides the platform with flexibility in regard to how to match.

%
%
%
%

\paragraph{\textbf{Case 4: platform matching under fixed fees.}}
As an extension of imposing caps on all fees, we consider a regulatory policy where the platform is required to keep the same fee structure as it picks, optimally, in a world without an economic shock, but the platform may adapt its matching policy in response to shocks.
This is of interest because, unlike platform fees which are observable, matching algorithms used by a platform are often proprietary and hard to regulate, bringing additional challenges in aligning platform incentives with broader societal considerations.
%

\begin{wrapfigure}[15]{r}{0.42\textwidth}
    \centering
    \includegraphics[width=0.36\textwidth]{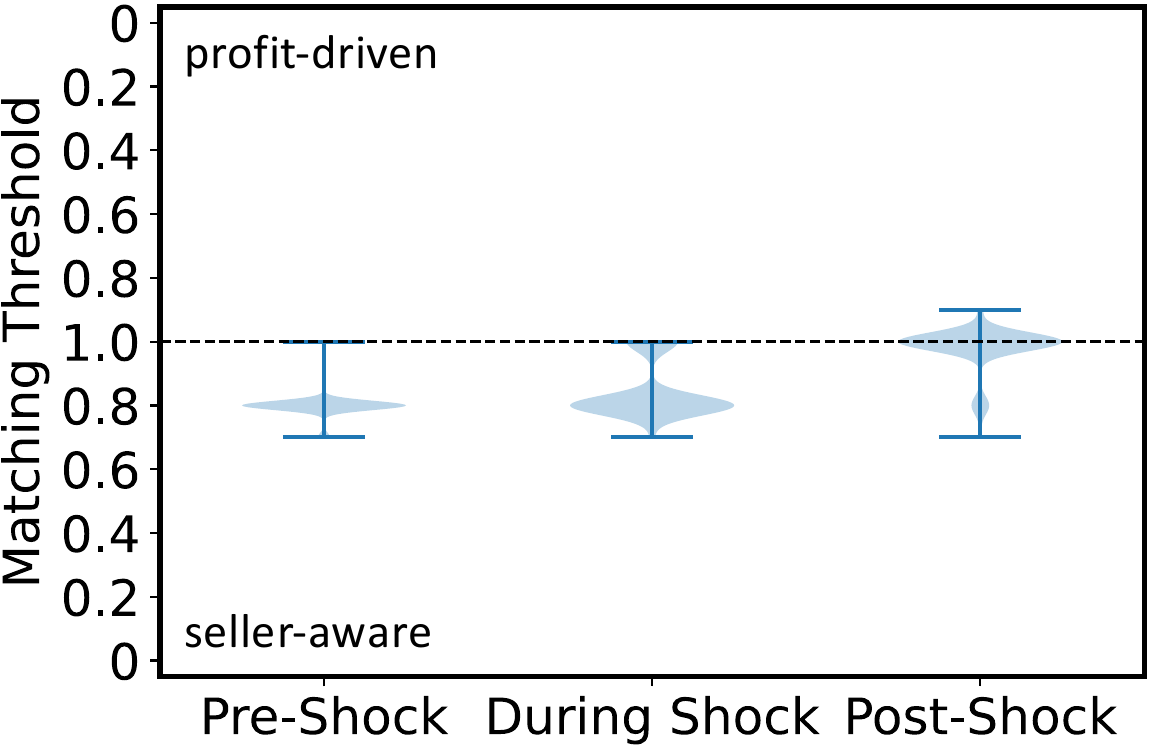}
    \caption[width=0.36\textwidth]{Probability density of the matching rule and the matching utility threshold chosen by the platform in the pre-, during, and post-shock stages. 
    \label{fig:matching_thresh}}
\end{wrapfigure}

In particular, we follow Section~\ref{sec:matching_policy} and consider a platform that uses RL to learn a matching policy under the fixed, optimal fees (i.e., $P_{\B, k} = 1.2$, $P_{\S, k}=2.0$, and $P_{R, k}=0.1$) that are chosen by a revenue-optimizing platform based on the use of Bayesian optimization in the no-shock setting.
As shown in Figure~\ref{fig:all_obj_results}a, column 10, 
even together with strategic matching, the effect of the platform is  to help the system preserve a similar level of economic welfare post-shock and result in fewer bankrupt sellers (Figure~\ref{fig:all_obj_results}c, column 10).
%

%
Figure~\ref{fig:matching_thresh} visualizes the learned matching policy.
%
We can see that to maximize revenue, the platform learns to generally adopt the seller-aware rule, which  matches to sellers with lower on-platform surplus, both before and during the shock. 
This can be explained as follows: (1)~the regulated, relatively low fees motivate the platform to retain a larger number of sellers through matching, so as to generate more revenue from registration fees, and (2)~with the shock affecting world transactions, the platform has more flexibility in matching, and can afford to compromise a bit on matching quality during the shock without losing many buyers. 
As the shock decays, the platform tends to increase the matching utility threshold to provide buyers with better matches, and  sometimes makes use of the profit-driven rule to extract more revenue from transaction fees paid by high-price sellers.

%
%
To further understand the role of the platform, we compare the use of RL matching with \textit{myopic query matching} 
 (Figure~\ref{fig:all_obj_results}, column 9) under the same set of fixed fees.
Overall, as shown in Table~\ref{table:matching_stats}, the RL-matching platform achieves both higher revenue and total welfare compared to  myopic-matching; it also substantially reduces the probability of cheap and niche sellers going bankrupt due to the shocks. 
%

%

%
%

%

\begin{table}[t]
		\centering
            \small
		\begin{tabular}{llll}
			\textbf{\ \ } & \textbf{Myopic matching \ \ } & \textbf{RL matching} \\
			\hline\midrule
			Welfare & 973.12 (5.29) & 989.44 (4.40)\\[0.3em]
			Revenue & 241.63 (3.06) & 261.28 (1.59)\\[0.1em]
   \midrule
			Seller shutdown freq. & 0.44 (0.02) & 0.34 (0.01)\\[0.3em]
			Core seller shutdown freq. & 0.30 (0.01) & 0.21 (0.01)\\[0.3em]
			Niche seller shutdown freq. & 0.23 (0.02) & 0.10 (0.02)\\[0.3em]
			Cheap seller shutdown freq.$\ \ \ $ & 0.86 (0.02) & 0.78 (0.02)\\\hline\hline
		\end{tabular}
	\caption{Welfare, platform revenue, and seller bankrupt frequencies of markets mediated by a myopic-matching platform and a RL-matching platform.} 
	\label{table:matching_stats}
        \vspace{-1ex}
\end{table}

\section{Discussion}
\label{sec:extensions}

In this paper, we have introduced a multi-agent Gym environment with which to study
 a platform-mediated economy in the presence of market shocks, making use of RL to model the decision making of a rational, self-interested platform.
 This is an important new application of AI techniques given the prominence of platforms in today's economic systems and the complexity of these systems.
Through this framework, we study and interpret the effect of several regulatory considerations on the efficiency and resilience of the overall economic system under optimal platform responses, confirming caution in regard to some interventions and suggesting a new kind of intervention. 

There are several interesting extensions that can be made to the current framework, including platform policies that combine fee-setting and matching, platforms with incomplete information about agent queries and locations, allowing buyer preferences or seller offerings to adapt over time, allowing buyers to learn from on-platform matches about sellers to match with off platform, and achieving economic systems at larger scale. 

While we think the present approach holds great promise, 
we note that there  also remain a number of  challenges that  need to be addressed 
before  AI-based simulation frameworks can be used
to guide the design of real-world economic systems. 
For example, 
 we need principled ways, together with data collection from real-world, platform-based markets, 
 to better inform modeling choices and calibrate the developed simulation framework.
It is also  important to interpret the learned policies  in order
to understand such factors
as the fairness implications of platform behavior on all parties and
in providing transparency to all stakeholders. 
We leave these aspects for future research, with the goal
 that the multi-agent platform model developed here offers a constructive
basis for researchers and policymakers to
better evaluate regulatory interventions and understand their interplay with different parties of agents.

\bibliographystyle{ACM-Reference-Format}
\bibliography{refs}


\begin{thebibliography}{35}


\ifx \showCODEN    \undefined \def \showCODEN     #1{\unskip}     \fi
\ifx \showDOI      \undefined \def \showDOI       #1{#1}\fi
\ifx \showISBNx    \undefined \def \showISBNx     #1{\unskip}     \fi
\ifx \showISBNxiii \undefined \def \showISBNxiii  #1{\unskip}     \fi
\ifx \showISSN     \undefined \def \showISSN      #1{\unskip}     \fi
\ifx \showLCCN     \undefined \def \showLCCN      #1{\unskip}     \fi
\ifx \shownote     \undefined \def \shownote      #1{#1}          \fi
\ifx \showarticletitle \undefined \def \showarticletitle #1{#1}   \fi
\ifx \showURL      \undefined \def \showURL       {\relax}        \fi
\providecommand\bibfield[2]{#2}
\providecommand\bibinfo[2]{#2}
\providecommand\natexlab[1]{#1}
\providecommand\showeprint[2][]{arXiv:#2}

\bibitem[Ahuja et~al\mbox{.}(2021)]%
        {mckinsey2021}
\bibfield{author}{\bibinfo{person}{Kabir Ahuja}, \bibinfo{person}{Vishwa
  Chandra}, \bibinfo{person}{Victoria Lord}, {and} \bibinfo{person}{Curtis
  Peens}.} \bibinfo{year}{2021}\natexlab{}.
\newblock \showarticletitle{Ordering in: The rapid evolution of food delivery}.
\newblock \bibinfo{journal}{\emph{McKinsey \& Company}} (\bibinfo{year}{2021}).
\newblock


\bibitem[Armstrong(2006)]%
        {Armstrong2006}
\bibfield{author}{\bibinfo{person}{Mark Armstrong}.}
  \bibinfo{year}{2006}\natexlab{}.
\newblock \showarticletitle{Competition in Two-Sided Markets}.
\newblock \bibinfo{journal}{\emph{The RAND Journal of Economics}}
  \bibinfo{volume}{37}, \bibinfo{number}{3} (\bibinfo{year}{2006}),
  \bibinfo{pages}{668--691}.
\newblock
\showISSN{07416261}
\urldef\tempurl%
\url{http://www.jstor.org/stable/25046266}
\showURL{%
\tempurl}


\bibitem[Brero et~al\mbox{.}(2021)]%
        {BreroEGPR21}
\bibfield{author}{\bibinfo{person}{Gianluca Brero}, \bibinfo{person}{Alon
  Eden}, \bibinfo{person}{Matthias Gerstgrasser}, \bibinfo{person}{David~C.
  Parkes}, {and} \bibinfo{person}{Duncan Rheingans{-}Yoo}.}
  \bibinfo{year}{2021}\natexlab{}.
\newblock \showarticletitle{Reinforcement Learning of Sequential Price
  Mechanisms}. In \bibinfo{booktitle}{\emph{Thirty-Fifth {AAAI} Conference on
  Artificial Intelligence}}. \bibinfo{pages}{5219--5227}.
\newblock


\bibitem[Caillaud and Jullien(2003)]%
        {Caillaud2003}
\bibfield{author}{\bibinfo{person}{Bernard Caillaud} {and}
  \bibinfo{person}{Bruno Jullien}.} \bibinfo{year}{2003}\natexlab{}.
\newblock \showarticletitle{Chicken \& Egg: Competition among Intermediation
  Service Providers}.
\newblock \bibinfo{journal}{\emph{The RAND Journal of Economics}}
  \bibinfo{volume}{34}, \bibinfo{number}{2} (\bibinfo{year}{2003}),
  \bibinfo{pages}{309--328}.
\newblock
\showISSN{07416261}
\urldef\tempurl%
\url{http://www.jstor.org/stable/1593720}
\showURL{%
\tempurl}


\bibitem[Chen et~al\mbox{.}(2019)]%
        {Chen2019}
\bibfield{author}{\bibinfo{person}{Minmin Chen}, \bibinfo{person}{Alex Beutel},
  \bibinfo{person}{Paul Covington}, \bibinfo{person}{Sagar Jain},
  \bibinfo{person}{Francois Belletti}, {and} \bibinfo{person}{Ed Chi}.}
  \bibinfo{year}{2019}\natexlab{}.
\newblock \showarticletitle{Top-K Off-Policy Correction for a REINFORCE
  Recommender System}. In \bibinfo{booktitle}{\emph{Proceedings of the 12th ACM
  International Conference on Web Search and Data Mining.}}
  \bibinfo{pages}{456--464}.
\newblock


\bibitem[Degris et~al\mbox{.}(2012)]%
        {Degris2012}
\bibfield{author}{\bibinfo{person}{Thomas Degris}, \bibinfo{person}{Patrick~M.
  Pilarski}, {and} \bibinfo{person}{Richard~S. Sutton}.}
  \bibinfo{year}{2012}\natexlab{}.
\newblock \showarticletitle{Model-Free reinforcement learning with continuous
  action in practice}. In \bibinfo{booktitle}{\emph{2012 American Control
  Conference (ACC)}}. \bibinfo{pages}{2177--2182}.
\newblock


\bibitem[Dub{\'e} et~al\mbox{.}(2009)]%
        {dube2009switching}
\bibfield{author}{\bibinfo{person}{Jean-Pierre Dub{\'e}},
  \bibinfo{person}{G{\"u}nter~J Hitsch}, {and} \bibinfo{person}{Peter~E
  Rossi}.} \bibinfo{year}{2009}\natexlab{}.
\newblock \showarticletitle{Do switching costs make markets less competitive?}
\newblock \bibinfo{journal}{\emph{Journal of Marketing research}}
  \bibinfo{volume}{46}, \bibinfo{number}{4} (\bibinfo{year}{2009}),
  \bibinfo{pages}{435--445}.
\newblock


\bibitem[Dub{\'e} et~al\mbox{.}(2010)]%
        {dube2010state}
\bibfield{author}{\bibinfo{person}{Jean-Pierre Dub{\'e}},
  \bibinfo{person}{G{\"u}nter~J Hitsch}, {and} \bibinfo{person}{Peter~E
  Rossi}.} \bibinfo{year}{2010}\natexlab{}.
\newblock \showarticletitle{State dependence and alternative explanations for
  consumer inertia}.
\newblock \bibinfo{journal}{\emph{The RAND Journal of Economics}}
  \bibinfo{volume}{41}, \bibinfo{number}{3} (\bibinfo{year}{2010}),
  \bibinfo{pages}{417--445}.
\newblock


\bibitem[Farrell and Shapiro(1988)]%
        {farrell1988dynamic}
\bibfield{author}{\bibinfo{person}{Joseph Farrell} {and} \bibinfo{person}{Carl
  Shapiro}.} \bibinfo{year}{1988}\natexlab{}.
\newblock \showarticletitle{Dynamic competition with switching costs}.
\newblock \bibinfo{journal}{\emph{The RAND Journal of Economics}}
  (\bibinfo{year}{1988}), \bibinfo{pages}{123--137}.
\newblock


\bibitem[Handel(2013)]%
        {handel2013adverse}
\bibfield{author}{\bibinfo{person}{Benjamin~R Handel}.}
  \bibinfo{year}{2013}\natexlab{}.
\newblock \showarticletitle{Adverse selection and inertia in health insurance
  markets: When nudging hurts}.
\newblock \bibinfo{journal}{\emph{American Economic Review}}
  \bibinfo{volume}{103}, \bibinfo{number}{7} (\bibinfo{year}{2013}),
  \bibinfo{pages}{2643--82}.
\newblock


\bibitem[Hausknecht and Stone(2015)]%
        {Hausknecht2015}
\bibfield{author}{\bibinfo{person}{Matthew~J. Hausknecht} {and}
  \bibinfo{person}{Peter Stone}.} \bibinfo{year}{2015}\natexlab{}.
\newblock \showarticletitle{Deep Recurrent Q-Learning for Partially Observable
  MDPs}.
\newblock \bibinfo{journal}{\emph{CoRR}}  \bibinfo{volume}{abs/1507.06527}
  (\bibinfo{year}{2015}).
\newblock


\bibitem[Heess et~al\mbox{.}(2015)]%
        {Heess2015}
\bibfield{author}{\bibinfo{person}{Nicolas Heess}, \bibinfo{person}{Jonathan~J.
  Hunt}, \bibinfo{person}{Timothy~P. Lillicrap}, {and} \bibinfo{person}{David
  Silver}.} \bibinfo{year}{2015}\natexlab{}.
\newblock \showarticletitle{Memory-based control with recurrent neural
  networks}.
\newblock \bibinfo{journal}{\emph{CoRR}}  \bibinfo{volume}{abs/1512.04455}
  (\bibinfo{year}{2015}).
\newblock


\bibitem[Honka(2014)]%
        {honka2014quantifying}
\bibfield{author}{\bibinfo{person}{Elisabeth Honka}.}
  \bibinfo{year}{2014}\natexlab{}.
\newblock \showarticletitle{Quantifying search and switching costs in the US
  auto insurance industry}.
\newblock \bibinfo{journal}{\emph{The RAND Journal of Economics}}
  \bibinfo{volume}{45}, \bibinfo{number}{4} (\bibinfo{year}{2014}),
  \bibinfo{pages}{847--884}.
\newblock


\bibitem[Ie et~al\mbox{.}(2019)]%
        {Ie2019}
\bibfield{author}{\bibinfo{person}{Eugene Ie}, \bibinfo{person}{Chih{-}Wei
  Hsu}, \bibinfo{person}{Martin Mladenov}, \bibinfo{person}{Vihan Jain},
  \bibinfo{person}{Sanmit Narvekar}, \bibinfo{person}{Jing Wang},
  \bibinfo{person}{Rui Wu}, {and} \bibinfo{person}{Craig Boutilier}.}
  \bibinfo{year}{2019}\natexlab{}.
\newblock \showarticletitle{RecSim: {A} Configurable Simulation Platform for
  Recommender Systems}.
\newblock \bibinfo{journal}{\emph{CoRR}}  \bibinfo{volume}{abs/1909.04847}
  (\bibinfo{year}{2019}).
\newblock


\bibitem[Kaelbling et~al\mbox{.}(1998)]%
        {KAELBLING1998}
\bibfield{author}{\bibinfo{person}{Leslie~Pack Kaelbling},
  \bibinfo{person}{Michael~L. Littman}, {and} \bibinfo{person}{Anthony~R.
  Cassandra}.} \bibinfo{year}{1998}\natexlab{}.
\newblock \showarticletitle{Planning and acting in partially observable
  stochastic domains}.
\newblock \bibinfo{journal}{\emph{Artificial Intelligence}}
  \bibinfo{volume}{101}, \bibinfo{number}{1} (\bibinfo{year}{1998}),
  \bibinfo{pages}{99--134}.
\newblock
\showISSN{0004-3702}


\bibitem[Li and Wang(2021)]%
        {li2021}
\bibfield{author}{\bibinfo{person}{Zhuoxin Li} {and} \bibinfo{person}{Gang
  Wang}.} \bibinfo{year}{2021}\natexlab{}.
\newblock \bibinfo{title}{Regulating Powerful Platforms: Evidence from
  Commission Fee Caps in On-Demand Services}.
\newblock
\newblock
\urldef\tempurl%
\url{https://ssrn.com/abstract=3871514}
\showURL{%
\tempurl}


\bibitem[MacKay and Remer(2021)]%
        {MacKay2021}
\bibfield{author}{\bibinfo{person}{Alexander MacKay} {and}
  \bibinfo{person}{Marc Remer}.} \bibinfo{year}{2021}\natexlab{}.
\newblock \bibinfo{title}{Consumer Inertia and Market Power}.
\newblock
\newblock
\urldef\tempurl%
\url{https://ssrn.com/abstract=3380390}
\showURL{%
\tempurl}


\bibitem[Mladenov et~al\mbox{.}(2020)]%
        {Mladenov2020}
\bibfield{author}{\bibinfo{person}{Martin Mladenov}, \bibinfo{person}{Elliot
  Creager}, \bibinfo{person}{Omer Ben{-}Porat}, \bibinfo{person}{Kevin
  Swersky}, \bibinfo{person}{Richard~S. Zemel}, {and} \bibinfo{person}{Craig
  Boutilier}.} \bibinfo{year}{2020}\natexlab{}.
\newblock \showarticletitle{Optimizing Long-term Social Welfare in Recommender
  Systems: {A} Constrained Matching Approach}. In
  \bibinfo{booktitle}{\emph{Proceedings of the 27th International Conference on
  Machine Learning}}. \bibinfo{pages}{6987--6998}.
\newblock


\bibitem[Mnih et~al\mbox{.}(2016)]%
        {Mnih2016}
\bibfield{author}{\bibinfo{person}{Volodymyr Mnih},
  \bibinfo{person}{Adria~Puigdomenech Badia}, \bibinfo{person}{Mehdi Mirza},
  \bibinfo{person}{Alex Graves}, \bibinfo{person}{Timothy Lillicrap},
  \bibinfo{person}{Tim Harley}, \bibinfo{person}{David Silver}, {and}
  \bibinfo{person}{Koray Kavukcuoglu}.} \bibinfo{year}{2016}\natexlab{}.
\newblock \showarticletitle{Asynchronous Methods for Deep Reinforcement
  Learning}. In \bibinfo{booktitle}{\emph{Proceedings of The 33rd International
  Conference on Machine Learning}}, Vol.~\bibinfo{volume}{48}.
  \bibinfo{pages}{1928--1937}.
\newblock


\bibitem[{National Restaurant Association}(2020)]%
        {NRA2020}
\bibfield{author}{\bibinfo{person}{{National Restaurant Association}}.}
  \bibinfo{year}{2020}\natexlab{}.
\newblock \showarticletitle{Restaurant Industry in Free Fall; 10,000 Close in
  Three Months}.
\newblock  (\bibinfo{year}{2020}).
\newblock


\bibitem[Nogueira(2014)]%
        {bo}
\bibfield{author}{\bibinfo{person}{Fernando Nogueira}.}
  \bibinfo{year}{2014}\natexlab{}.
\newblock \bibinfo{title}{{Bayesian Optimization}: Open source constrained
  global optimization tool for {Python}}.
\newblock
\newblock
\urldef\tempurl%
\url{https://github.com/fmfn/BayesianOptimization}
\showURL{%
\tempurl}


\bibitem[Oblander and McCarthy(2022)]%
        {oblander2022}
\bibfield{author}{\bibinfo{person}{E.~Shin Oblander} {and}
  \bibinfo{person}{Daniel McCarthy}.} \bibinfo{year}{2022}\natexlab{}.
\newblock \bibinfo{title}{Persistence of Consumer Lifestyle Choices: Evidence
  from Restaurant Delivery During COVID-19}.
\newblock
\newblock
\urldef\tempurl%
\url{https://ssrn.com/abstract=3836262}
\showURL{%
\tempurl}


\bibitem[Papadimitriou et~al\mbox{.}(2021)]%
        {pdp}
\bibfield{author}{\bibinfo{person}{Christos Papadimitriou},
  \bibinfo{person}{Kiran Vodrahalli}, {and} \bibinfo{person}{Mihalis
  Yannakakis}.} \bibinfo{year}{2021}\natexlab{}.
\newblock \showarticletitle{The Platform Design Problem}. In
  \bibinfo{booktitle}{\emph{Web and Internet Economics (WINE)}}.
  \bibinfo{pages}{317--333}.
\newblock


\bibitem[Papadimitriou et~al\mbox{.}(2016)]%
        {PapadimitriouPP16}
\bibfield{author}{\bibinfo{person}{Christos~H. Papadimitriou},
  \bibinfo{person}{George Pierrakos}, \bibinfo{person}{Christos{-}Alexandros
  Psomas}, {and} \bibinfo{person}{Aviad Rubinstein}.}
  \bibinfo{year}{2016}\natexlab{}.
\newblock \showarticletitle{On the Complexity of Dynamic Mechanism Design}. In
  \bibinfo{booktitle}{\emph{Proceedings of the Twenty-Seventh Annual {ACM-SIAM}
  Symposium on Discrete Algorithms}}, \bibfield{editor}{\bibinfo{person}{Robert
  Krauthgamer}} (Ed.). \bibinfo{publisher}{{SIAM}},
  \bibinfo{pages}{1458--1475}.
\newblock


\bibitem[Raj et~al\mbox{.}(2021)]%
        {raj2021}
\bibfield{author}{\bibinfo{person}{Manav Raj}, \bibinfo{person}{Arun
  Sundararajan}, {and} \bibinfo{person}{Calum You}.}
  \bibinfo{year}{2021}\natexlab{}.
\newblock \bibinfo{title}{COVID-19 and Digital Resilience: Evidence from Uber
  Eats}.
\newblock
\newblock
\urldef\tempurl%
\url{https://ssrn.com/abstract=3625638}
\showURL{%
\tempurl}


\bibitem[Rochet and Tirole(2003)]%
        {Rochet2003}
\bibfield{author}{\bibinfo{person}{Jean-Charles Rochet} {and}
  \bibinfo{person}{Jean Tirole}.} \bibinfo{year}{2003}\natexlab{}.
\newblock \showarticletitle{{Platform Competition in Two-Sided Markets}}.
\newblock \bibinfo{journal}{\emph{Journal of the European Economic
  Association}} \bibinfo{volume}{1}, \bibinfo{number}{4} (\bibinfo{date}{06}
  \bibinfo{year}{2003}), \bibinfo{pages}{990--1029}.
\newblock
\showISSN{1542-4766}


\bibitem[Salakhutdinov and Mnih(2007)]%
        {Salakhutdinov2007}
\bibfield{author}{\bibinfo{person}{Ruslan Salakhutdinov} {and}
  \bibinfo{person}{Andriy Mnih}.} \bibinfo{year}{2007}\natexlab{}.
\newblock \showarticletitle{Probabilistic Matrix Factorization}. In
  \bibinfo{booktitle}{\emph{Proceedings of the 20th International Conference on
  Neural Information Processing Systems}}. \bibinfo{pages}{1257--1264}.
\newblock


\bibitem[Shen et~al\mbox{.}(2020)]%
        {ShenPLZQHGDLT20}
\bibfield{author}{\bibinfo{person}{Weiran Shen}, \bibinfo{person}{Binghui
  Peng}, \bibinfo{person}{Hanpeng Liu}, \bibinfo{person}{Michael Zhang},
  \bibinfo{person}{Ruohan Qian}, \bibinfo{person}{Yan Hong},
  \bibinfo{person}{Zhi Guo}, \bibinfo{person}{Zongyao Ding},
  \bibinfo{person}{Pengjun Lu}, {and} \bibinfo{person}{Pingzhong Tang}.}
  \bibinfo{year}{2020}\natexlab{}.
\newblock \showarticletitle{Reinforcement Mechanism Design: With Applications
  to Dynamic Pricing in Sponsored Search Auctions}. In
  \bibinfo{booktitle}{\emph{The Thirty-Fourth {AAAI} Conference on Artificial
  Intelligence}}. \bibinfo{publisher}{{AAAI} Press},
  \bibinfo{pages}{2236--2243}.
\newblock


\bibitem[Shum(2004)]%
        {shum2004does}
\bibfield{author}{\bibinfo{person}{Matthew Shum}.}
  \bibinfo{year}{2004}\natexlab{}.
\newblock \showarticletitle{Does advertising overcome brand loyalty? Evidence
  from the breakfast-cereals market}.
\newblock \bibinfo{journal}{\emph{Journal of Economics \& Management Strategy}}
  \bibinfo{volume}{13}, \bibinfo{number}{2} (\bibinfo{year}{2004}),
  \bibinfo{pages}{241--272}.
\newblock


\bibitem[Sutton and Barto(1998)]%
        {rlbook}
\bibfield{author}{\bibinfo{person}{Richard~S. Sutton} {and}
  \bibinfo{person}{Andrew~G. Barto}.} \bibinfo{year}{1998}\natexlab{}.
\newblock \bibinfo{booktitle}{\emph{Reinforcement Learning: an Introduction}}.
\newblock \bibinfo{publisher}{MIT Press}.
\newblock


\bibitem[Tang(2017)]%
        {Tang17abc}
\bibfield{author}{\bibinfo{person}{Pingzhong Tang}.}
  \bibinfo{year}{2017}\natexlab{}.
\newblock \showarticletitle{Reinforcement mechanism design}. In
  \bibinfo{booktitle}{\emph{Proceedings of the Twenty-Sixth International Joint
  Conference on Artificial Intelligence}}. \bibinfo{pages}{5146--5150}.
\newblock


\bibitem[Wierstra et~al\mbox{.}(2007)]%
        {Wierstra2007}
\bibfield{author}{\bibinfo{person}{Daan Wierstra}, \bibinfo{person}{Alexander
  Foerster}, \bibinfo{person}{Jan Peters}, {and} \bibinfo{person}{J{\"u}rgen
  Schmidhuber}.} \bibinfo{year}{2007}\natexlab{}.
\newblock \showarticletitle{Solving Deep Memory POMDPs with Recurrent Policy
  Gradients}. In \bibinfo{booktitle}{\emph{Artificial Neural Networks -- ICANN
  2007}}, \bibfield{editor}{\bibinfo{person}{Joaquim~Marques de~S{\'a}},
  \bibinfo{person}{Lu{\'i}s~A. Alexandre}, \bibinfo{person}{W{\l}odzis{\l}aw
  Duch}, {and} \bibinfo{person}{Danilo Mandic}} (Eds.).
  \bibinfo{pages}{697--706}.
\newblock


\bibitem[Williams and Peng(1991)]%
        {Williams1991}
\bibfield{author}{\bibinfo{person}{Ronald~J. Williams} {and}
  \bibinfo{person}{Jing Peng}.} \bibinfo{year}{1991}\natexlab{}.
\newblock \showarticletitle{Function Optimization using Connectionist
  Reinforcement Learning Algorithms}.
\newblock \bibinfo{journal}{\emph{Connection Science}}  \bibinfo{volume}{3}
  (\bibinfo{year}{1991}), \bibinfo{pages}{241--268}.
\newblock


\bibitem[Zhan et~al\mbox{.}(2021)]%
        {Zhan2021}
\bibfield{author}{\bibinfo{person}{Ruohan Zhan}, \bibinfo{person}{Konstantina
  Christakopoulou}, \bibinfo{person}{Ya Le}, \bibinfo{person}{Jayden Ooi},
  \bibinfo{person}{Martin Mladenov}, \bibinfo{person}{Alex Beutel},
  \bibinfo{person}{Craig Boutilier}, \bibinfo{person}{Ed Chi}, {and}
  \bibinfo{person}{Minmin Chen}.} \bibinfo{year}{2021}\natexlab{}.
\newblock \showarticletitle{Towards Content Provider Aware Recommender Systems:
  A Simulation Study on the Interplay between User and Provider Utilities}. In
  \bibinfo{booktitle}{\emph{Proceedings of the Web Conference 2021}}.
  \bibinfo{pages}{3872–3883}.
\newblock


\bibitem[Zheng et~al\mbox{.}(2022)]%
        {Zheng2021}
\bibfield{author}{\bibinfo{person}{Stephan Zheng}, \bibinfo{person}{Alexander
  Trott}, \bibinfo{person}{Sunil Srinivasa}, \bibinfo{person}{David~C. Parkes},
  {and} \bibinfo{person}{Richard Socher}.} \bibinfo{year}{2022}\natexlab{}.
\newblock \showarticletitle{The {AI} Economist: Taxation policy design via
  two-level deep multiagent reinforcement learning}.
\newblock \bibinfo{journal}{\emph{Science Advances}} \bibinfo{volume}{8},
  \bibinfo{number}{18} (\bibinfo{year}{2022}), \bibinfo{pages}{eabk2607}.
\newblock


\end{thebibliography}

\appendix
\section{Deferred Materials from Section~\ref{sec:example}} \label{sec:apx-example}
\subsection{Stackelberg Equilibrium Definition for Section~\ref{sec:example}}
The {\em Stackelberg equilibrium} for the platform economy in Section~\ref{sec:example} is:
\begin{enumerate}
    \item The platform first decides upon a buyers' fee $P_\B$ and a sellers' fee $P_\S$. 
    If the platform adopts non-myopic matching, it also decides upon a matching policy.
    \item Given the platform's fees and matching policy, buyers and sellers decide to join the platform by comparing surpluses achieved on- and off-platform.
    That is, agents are playing an equilibrium where no agent can improve their surplus by making a different decision in regard to joining the platform.
    \item The platform can not increase its objective (e.g., revenue) by changing their fees or matching policy and allowing the agents to adapt their behavior.
\end{enumerate}
\subsection{Proofs of Section \ref{sec:example}} \label{sec:apx-examples}

\begin{proof}[Proof of Case 1]
    When there is no platform, Buyer 1 does not know Seller 1 and can only match their queries to Seller 2. 
    Because of the world transaction friction $\mu=1$, matching queries of type $Q_{11}$ to Seller 2 provides the buyer with negative surplus, i.e., $r(Q_{11},S_2, 1)= 2-|Q_{11}-S_2|-1 \cdot \mu = \epsilon-1<0$.
    Therefore, Buyer 1 will only match queries of type $Q_{12}$, giving a total surplus of 
    $$r(Q_{12},S_2, 1) \cdot m = (2-|Q_{12}-S_2|-1) \cdot m = \epsilon m.$$
    Similarly, with the high friction, Buyer 2 will get negative surplus from matching its queries to either seller, and thus will choose not to transact. 
    As for the sellers, Seller 1 does not get any transaction to cover its fixed cost, and has a negative surplus of $-m$, causing the seller to go bankrupt. 
    Seller 2 gets to transact with $m$ queries, obtaining a zero surplus.
    The total welfare of the economic system is $\epsilon m$.
\end{proof}
\begin{proof}[Proof of Case 2]
    With myopic optimal matching, the platform reasons about which agents to keep on the platform, such that in equilibrium, the maximal fees it can set maximizes revenue while retaining the agents to be on-platform. 
    We consider the following options:
    \begin{enumerate}
        \item \textit{All agents join the platform.} In this case, the myopic matching ensures that  $Q_{11}$-type queries are matched to Seller 1, and $Q_{12}$- and $Q_2$-type queries are matched to Seller 2 via the platform, i.e., $\I_w = 0$. 
        Since Seller 1 gets zero surplus considering the fixed cost of $m$, the platform must set $P_\S=0$ in order to keep this seller on-platform. 
        Assuming no buyer registration fee, Buyer 1's surplus is 
        $$r(Q_{11},S_1, 0) \cdot m + r(Q_{12},S_2, 0) \cdot m = (2-|Q_{11}-S_1|) \cdot m + (2-|Q_{12}-S_2|) \cdot m = 3m,$$
        whereas with no platform (Case 1), this buyer has surplus $\epsilon m$. 
        Similarly, Buyer 2's surplus is $r(Q_{2},S_2, 0) \cdot m = (2-2\epsilon)m$, while without a platform, its surplus is zero. 
        Therefore, the platform can set a maximal fee $P_\B=\min\{3m-\epsilon m, (2-2\epsilon)m\}=(2-2\epsilon)m$ in order to keep both buyers on-platform. The platform revenue in this case is $2P_\B + 2P_\S=(4-4\epsilon)m$.
        \item \textit{All but Seller 1 join the platform.} The myopic policy matches all $4m$ queries to Seller 2 (recall that Buyer 1 cannot transact with Seller 1 without the platform's recommendation). 
        Without fees, Seller 2 has a surplus of $3m$, compared with a surplus of zero without joining, so the platform can set $P_\S=3m$. 
        Since Seller 1 receives no query, it goes bankrupt. 
        Without fees, Buyer 1's surplus on platform is
        $$r(Q_{11},S_2, 0) \cdot m + r(Q_{12},S_2, 0) \cdot m = \epsilon m+(1+\epsilon) m = (1+2\epsilon) m,$$ 
        compared with $\epsilon m$ from not joining.
        Similarly, Buyer 2's surplus is $(2-2\epsilon)m$, while the off-platform surplus is zero. 
        Therefore, to have both buyers join the platform, the platform can set $P_\B=\min\{(1+2\epsilon) m-\epsilon m, (2-2\epsilon)m\}=(1+\epsilon) m$ with infinitesimal $\epsilon$.
        This give a total revenue of $2P_\B + P_\S = 2(1+\epsilon) m + 3m = (5+2\epsilon)m.$   
        
        \item \textit{Seller 2 does not join the platform.} In this case, Buyer 2 cannot get a positive surplus either on- or off-platform, and will also stay off the platform. 
        Without fees, Buyer 1's surplus on platform is $$r(Q_{11},S_1, 0) \cdot m + r(Q_{12},S_1, 0) \cdot m = (2-\epsilon)m+(1-\epsilon) m = (3-2\epsilon)m,$$ compared with $\epsilon m$ off  platform, and the platform can set $P_\B=(3-3\epsilon)m$.
       Without fees, Seller 1's surplus is $m$, being matched to $2m$ queries, while its off-platform surplus is $0$. 
       The platform sets $P_\S=m$, and in this case, the platform revenue is $P_\B+P_\S=(4-3\epsilon)m$.
    \end{enumerate}
    
    Comparing all options, a revenue-maximizing platform will choose the second one, with all agents but Seller 1 joining the platform.
\end{proof}
\begin{proof}[Proof of Case 3]
    In this case, the platform needs to decide (1)~who joins the platform and (2)~how many queries of each type to match to which on-platform seller.
    We first note that 
    %
    if there is only one seller joining the platform, the platform's matching policy is necessarily myopic, which reduces to Case 2 and the maximum revenue is $(5+2\epsilon)m < 6m$. 
    Consider the case of a single buyer joining the platform. 
    Since there are only $2m$ queries to be matched, the sellers' total surplus is at most $m$ and the buyer's surplus is at most $4m$, therefore the revenue cannot be greater than $5m$. 
    
    We consider the case when all agents join the platform.
    Note that all queries of $Q_2$ must be matched to Seller 2, otherwise Buyer 2 will reject the matches. Given this, the platform aims to increase Seller 1's  surplus from matching, in order to increase the fee that can be charged to sellers. 
    To this end, the platform considers to divert $x<m$ queries of type $Q_{12}$ to Seller 1, with the rest $m-x$ matched to Seller 2.
    We note that even for $Q_{12}$-type queries, Buyer 1 prefers transacting on-platform with Seller 1 to transacting off-platform with Seller 2, due to the world transaction friction.
    Without fees, the relative surplus of each agents on- and off-platform is:
    \begin{itemize}
        \item \textit{Buyer 1.} On-platform: 
        $r(Q_{11},S_1, 0) \cdot m + r(Q_{12},S_1, 0) \cdot x + r(Q_{12},S_2, 0) \cdot (m-x) = 3m+2x$, 
        and off-platform: $\epsilon m$.
        \item \textit{Buyer 2.} On-platform: $(2-2\epsilon)m$, and off-platform: $0$.
        \item \textit{Seller 1.} On-platform: $m+x-m=x$, and off-platform: $0$.
        \item \textit{Seller 2.} On-platform: $2m+(m-x)-m=2m-x$, and off-platform: $0$.
    \end{itemize}
    Based on this, the platform can set $P_\B=\min\{3m+2x-\epsilon m, (2-2\epsilon)m\}=(2-2\epsilon)m$ to keep both buyers and $P_\S=\min\{x,2m-x\}=m$ under the optimal choice of $x=m$.
    The platform revenue is $2P_\B + 2P_\S = (6-4\epsilon)m$, with all agents joining the platform.
\end{proof}
\begin{proof}[Proof of Case 4]
    Consider a fee $P \geq 0$ for an on-platform agent, which is set to some value strictly less than the difference between their (without fee) on-platform surplus and off-platform surplus. 
    Increasing $P$ to $P+\delta$ for some infinitesimal $\delta>0$ increases the platform's objective by $(1-\alpha)\delta$, which is strictly larger than $0$ with $\alpha < 1$. 
    Thus, for whichever agents join the platform, the platform should set fees to maximize its own revenue (under the constraint that those agents do not leave). 
    Similar to Case 2, the platform again needs to consider different sets of agents who might join the platform. 
    We consider the following cases:  
    \begin{enumerate}
        \item \textit{All agents join the platform.} 
        To maximize its revenue, the platform sets fees $P_\B=(2-2\epsilon)m$ and $P_\S=0$ while keeping all agents on the platform.
        Under such fees, Buyer 1 has a per-epoch surplus of $r(Q_{11},S_1, 0) \cdot m + r(Q_{12},S_2, 0) \cdot m - P_\B = (1+2\epsilon) m$, and Buyer 2 has a per-epoch surplus of $r(Q_{2},S_2, 0) \cdot 2m -P_\B= 0$. 
        Seller 1 gets to match with $m$ queries and has a zero surplus, and Seller 2 gets to match with $3m$ queries and has an epoch surplus of $2m$.
        Therefore, the platform's objective value is $(4-4\epsilon)m + \alpha \cdot (3+2\epsilon) m.$
        \item \textit{All but Seller 1 join the platform.} 
        This is the same as Case 2.
        The platform sets $P_\B=(1+\epsilon)m$ and $P_\S=3m$. 
        The platform's objective value is $(5+2\epsilon)m +\alpha \cdot (1-2\epsilon)m.$
        \item \textit{Seller 2 does not join the platform.} 
        By the same scenario of Case 2, both the on-platform user surplus and the revenue are lower than the previous scenario where all but Seller 1 join the platform, and so is the objective value.
    \end{enumerate}
    
    Since $(4-4\epsilon)m + \alpha \cdot (3+2\epsilon) m > (5+2\epsilon)m +\alpha \cdot (1-2\epsilon)m$, whenever $\alpha > 1/2+ 2\epsilon/(1+2\epsilon)$, the best option is for all buyers and sellers to join the platform, leading no seller to go bankrupt and a total welfare of $(7-2\epsilon)m$. 
\end{proof}
\section{Subscription-Level Decision}
\label{app:model}

\label{app:counterfactual_estimates}



\paragraph{Calculations for Subscription Effect Estimation}
We first describe the subscription effect estimates for each type of the agents. 
To recap, these estimates, denoted $\xi_{k+1}$, are conducted based on the new platform fees and world transaction friction, under the same set of queries submitted or received in the last epoch, i.e., epoch $k$, and the same subscription decisions made by other agents.
%
%
\begin{itemize}[leftmargin=*, itemsep=0ex]
	\item \textbf{A buyer $b$, on-platform in epoch $k$}. 
	We denote the estimate of epoch surplus if buyer $b$ does not subscribe to the platform as $\xi^w_{b, k+1}$.
	For this, given each query of $b$  in epoch $k$, the choice in the world under new friction $\mu_{k+1}$ (Section~\ref{sec:choice_world_only}) is reevaluated
	\[ \xi^w_{b, k+1} = \sum_{t \in k} u^w_{b, t} (\mu_{k+1}) = \sum_{t \in k} \max\{u_{\B}(q_{b,t}, s^*_w) - \mu_{k+1}, 0\},\]
	where we denote $u^w_{b, t} (\mu)$ the world matching surplus for buyer $b$ at $t$ under the world friction $\mu$, i.e., $u^w_{b, t} (\mu) = \max\{u_{\B}(q_{b,t}, s^*_w) - \mu, 0\}$.
	Similarly, we denote $\xi^p_{b, k+1}$ the estimate of epoch surplus if buyer $b$ remains on platform.
	The new friction may affect the choice between a platform seller and a world seller. 
	This surplus is estimated based on updated decisions under $\mu_{k+1}$, with 
	\[ \xi^p_{b, k+1} = - P_{\B, k+1} + \sum_{t \in k} \max \{u^w_{b, t} (\mu_{k+1}), u^p_{b, t}\}.\]
	
	\item \textbf{A buyer $b$,  off-platform in epoch $k$}. 
	The surplus estimate of remaining off-platform depends on the new friction, i.e., 
	$\xi^w_{b, k+1} = \sum_{t \in k} u^w_{b, t} (\mu_{k+1})$.
	For $\xi^p_{b, k+1} $, we need to estimate the epoch surplus if $b$ subscribes to the platform.  We assume that buyer $b$ can observe platform-recommended sellers, e.g., this can be from trial periods to gain platform experience or a platform's estimate of costs and benefits based on past orders.
	For a query sequence $\{q_{b,t}\}_{t \in k}$,  denote the corresponding best platform-recommended sellers as $\{s^{p*}_{b,t}\}$.  The estimated surplus if subscribing to the platform is
	\[ \xi^p_{b, k+1} = - P_{\B, k+1} + \sum_{t \in k} \max \{u^w_{b, t} (\mu_{k+1}), u_{\B}(q_{b,t}, s^{p*}_{b,t})\}.\]

	\item \textbf{A  seller $s$, on-platform in epoch $k$}. 
	For $\xi^w_{s, k+1}$, we need to estimate the epoch surplus if $s$ does not subscribe to the platform, by reasoning about 
	(1)~how many more world transactions would happen if $s$ is not on the platform, and
	(2)~how buyers' transaction decisions may be affected by the epoch $k+1$ world friction. 
	To facilitate a precise estimate, we assume that seller $s$ can observe the sequence of query and seller candidate tuples $\{(q_{b, t}, s, s^p_{b,t})\}_{t\in k}$ in which seller $s$ is chosen as the best world option.
	Given $q_{b,t}$, $\S_k \wo s$, and a fixed platform matching strategy used in epoch $k$, we denote the updated, best-platform seller as $s^{p*}_{b,t}$.
	For this modified sequence $\{(q_{b, t}, s, s^{p*}_{b,t})\}$, we consider the choices buyers will make under the new friction $\mu_{k+1}$, and estimate the number of transactions  seller $s$ will receive without being on the platform, denoted $n^{w'}_{s, k}$.
	Thus, the estimated surplus if seller $s$  is off platform is 
	$\xi^w_{s, k+1} = n^{w'}_{s, k} v^1_s (1 - \omega_s)$.
	
	For $\xi^p_{s, k+1}$, we reason about how the new friction affects the number of transactions.
	Given the sequence $\{(q_{b, t}, s^w_{b,t}, s)\}_{t \in k}$ where $s$ is picked as the best platform seller, we consider  each buyer's choice and estimate the number of platform visits under $\mu_{k+1}$, denoted $n^{p^*}_{s, k}$.
	Given the sequence $\{(q_{b, t}, s, s^p_{b,t})\}$ where $s$ is picked as the best world seller, we estimate the number of world transactions under $\mu_{k+1}$ (even if seller $s$ chooses to stay on the platform), denoted $n^{w^*}_{s, k}$.
	Thus, the estimated  surplus that seller $s$ would receive  by subscribing to the platform is \[ \xi^p_{s, k+1} = - P_{\S, k+1} + n^{p^*}_{s, k} v^1_s (1 - \omega_s - P_{R, k+1}) + n^{w^*}_{s, k} v^1_s (1 - \omega_s).\]
	
	\item \textbf{A seller $s$, off-platform in epoch $k$}. 
	For $\xi^p_{s, k+1}$, we reason about the surplus from subscribing to the platform, e.g., by asking the platform for an estimate on the number of platform transactions. 
	Given the sequence of queries on the platform, i.e., $\{q_{b, t}\}_{b \in \B_k, t\in k}$, the platform can update the matches it would suggest $s$ also on-platform by following its matching strategy. 
	Denote the sequence of query and updated seller matches with tuples $\{(q_{b, t}, s^w_{b, t}, s^{p^*}_{b, t})\}_{b \in \B_k, t\in k}$. 
	With $\mu_{k+1}$, we estimate the  numbers of platform transactions $n^{p^*}_{s, k}$ and world transactions $n^{w^*}_{s, k}$.
	The estimated surplus if seller $s$  subscribes to the platform is 
	\[ \xi^p_{s, k+1} = - P_{\S, k+1} + n^{p^*}_{s, k} v^1_s (1 - \omega_s - P_{R, k+1}) + n^{w^*}_{s, k} v^1_s (1 - \omega_s).\]
	
	Seller $s$ also adjusts the surplus that they would get by remaining off platform, reasoning about how buyers' transaction decisions will be affected by the new world friction.
	Given the sequence of query and matched seller  tuples $\{(q_{b, t}, s, s^p_{b,t})\}_{b \in \B, t \in k}$ in which seller $s$ was chosen as the best world seller, we re-evaluate each buyer's choice to get $n^{w'}_{s, k}$.
	The estimated off-platform surplus is 
	\[ \xi^w_{s, k+1} = n^{w'}_{s, k} v^1_s (1 - \omega_s).\]
\end{itemize}

\paragraph{Calculations for Agent-Specific Decision Inertia} 
\label{sec:apx-inertia}

There is a rich body of literature that establishes \emph{decision inertia}, modeling the presence of such inertia across different markets (see Section~\ref{sec:choice_subscribe}). 
In our setting an agent, either a buyer or seller, in  subscription state $\I^p_k \in \{0, 1\}$, is prone to stay in the same state in epoch $k+1$, due to habit formation, loyalty, or inattention. 
We treat buyers and sellers in the same way and illustrate the concept with a buyer $b$  for simplicity.
%
%
Each buyer starts with an initial preference in regard to adopting the platform or not, denoted by an integer $\chi_{b,0} \sim U[-\chi, \chi]$ for some integer $\chi$. 
%
If the initial $\chi_{b,0}$ is positive, the buyer subscribes at the warm-up epoch, otherwise they do not. 
The  inertia $\chi_{b,k}$ maps into an additive bonus to either the surplus for joining the platform or remaining in the world through the functional form:  
\begin{equation}
	\sigma^p_{b, k+1} := \I^p_{b, k} \log \Parens{\chi_{b,k}}, 
	\qquad
	\sigma^w_{b, k+1} := (1-\I^p_{b, k}) \log \Parens{-\chi_{b,k}}.%
\end{equation}
That is, the longer an agent sticks to their decision, the larger the bias term gets, which increases in a concave way (logarithmically) over time.
Such interpretation of decision inertia as an additive bonus is common in the literature~\citep{MacKay2021,dube2009switching,farrell1988dynamic}. 
Based on this adjusted utility, agents  decide whether to subscribe or not according to probabilities inferred by the standard {\em discrete-choice logit} model~\citep{MacKay2021,dube2009switching}, where the probability of subscribing to the platform is
\begin{equation}
    \delta^p_{b, k+1} := \frac{\exp \Parens{\xi^p_{b,k+1} + \sigma^p_{b, k+1}}}{\exp \Parens{\xi^p_{b,k+1} + \sigma^p_{b, k+1}} + \exp \Parens{\xi^w_{b,k+1} + \sigma^w_{b, k+1}}}.
    \label{eq:discrete-choice-logit}
\end{equation}
%

The inertia is updated after each decision in the following way. 
If $\chi_{b,k}>0$ and the buyer subscribes, then $\chi_{b,k+1}:=\chi_{b,k}+1$, and otherwise $\chi_{b,k+1}:=-1$ (i.e., it resets for off-platform).
Similarly if $\chi_{b,k}<0$ and the buyer decides to stay off-platform, then $\chi_{b,k+1}:=\chi_{b,k}-1$, and otherwise reset $\chi_{b,k+1}:=1$. 

\section{The Platform's Matching Problem}
\label{app:pdp}
\subsection{Platform Matching Strategies}
\begin{algorithm}[H]
	\caption{The platform matching strategy with a matching rule and a utility threshold $\eta_k \in [0, 1]$.}
	\label{algo:greedy_matching}
	\begin{algorithmic}[1]
		\Statex \textbf{Input:}~%
		Platform fees $P_{\B, k}$, $P_{\S, k}$, and $P_{R, k}$. On-platform sellers $\S_k$ and their observable attributes.
		A buy query $q_{b,t}$ for $t \in k$ and $b \in \B_k$. 
		\Statex \textbf{Output:}~%
		A recommended seller $s^* \in \S_k$.
		\medskip
		\State Initialize a vector of estimated platform surplus for sellers $\vr^p_{\S, t}$ at the start of epoch $k$:
		\begin{equation*}
			r^p_{s,t} = 
			\begin{cases}
				-P_{\S, k} & \text{if $s \in \S_k$,} \\
				0 & \text{otherwise.}
			\end{cases}
		\end{equation*}
		\State Given $q_{b,t}$, compute $u^* \gets \max_{s \in \S_k} u_\B(q_{b,t}, s)$.
		\State $\S_u \gets \{s : s \in \S_k \text{ and } u_\B(q_{b,t}, s) \geq \eta_k u^*\}$
		\Comment{candidate sellers yield surplus at least $\eta_k u^*$.}
		\If {$\card{\S_u} = 1$}
		\State $s^* \gets  s \in \S_u$
		\Comment{the same seller as myopic matching.}
		\Else
		\If{\texttt{the seller-aware rule}}
		\State $\S_r \gets \{s : s \in \S_u \text{ and } r^p_{s,t} \leq 0\}$
		\If{$\S_r = \emptyset$} 
		\Comment{every candidate has break-even.}
		\State $r^p_{\min} \gets \min_{s \in \S_u} r^p_{s,t}$, $\S^* \gets \{s : s \in \S_u \text{ and } r^p_{s,t} = r^p_{\min}\}$
		\State $s^* \gets \argmax_{s \in \S^*} u_\B(q_{b,t}, s)$
		\Comment{the best candidate with lowest platform surplus.}
		\Else
		\State $r^p_{\max} \gets \max_{s \in \S_r} r^p_{s,t}$, $\S^* \gets \{s : s \in \S_r \text{ and } r^p_{s,t} = r^p_{\max}\}$
		\State $s^* \gets \argmax_{s \in \S^*} u_\B(q_{b,t}, s)$
		\Comment{the best candidate who is closest to break-even.}
		\EndIf	
		\Else
		\State $s^* \gets \argmax_{s \in \S^*} P_{R,k} v^1_s$
		\Comment{\texttt{the profit-driven rule}}
		\EndIf
		\EndIf
		\State $r^p_{s^*, t} \gets r^p_{s^*, t} + v^1_{s^*}(1-\omega_{s^*} -P_{R, k})$
		\State \Return $s^*$
	\end{algorithmic}
\end{algorithm}

\subsection{The Platform Matching POMDP}
\label{app:matching_policy_app}
Based on the fee-setting POMDP (Section~\ref{sec:pricing_policy}), we make the following adjustments to define the matching POMDP:
\begin{itemize} [leftmargin=*, itemsep=0ex]
	\item We define $x_k \in \X$ for epoch $k$ as the state of the market after the platform has set fees and agents have chosen to subscribe, but still before the first query is submitted.  
	That is, the state includes agent-subscription states $\I^p_{\B, k}$ and $\I^p_{\S, k}$ and  platform fees, i.e., $P_{\B, k}, P_{\S, k}, P_{R, k}$, for the upcoming epoch $k$.
	We also include the matching utility threshold adopted in the previous epoch, $\eta_{k-1}$, to state $x_k$.  
	All other elements of the state remain the same.
	\item Here, the platform's action $a_k$ chooses (i) the matching utility threshold for epoch $k$, and (ii) the matching rule for the epoch, whether seller-aware or profit-driven.
	For the threshold, we consider $\eta_k$ that takes discrete values in $[0,1]$.
	\item Different from the fee-setting POMDP, the state transition for platform matching starts with buyer queries: (1)~a buy agent generates a query, (2)~if the buyer is on platform, the platform recommends a seller based on $\eta_k$ and its matching rule, and (3)~the buyer selects a seller with whom to transact (Section~\ref{sec:choice_transact}).
	This gives the full sequence of queries, $Q_k$.
	At the end of epoch $k$, buyers and sellers observe the new fees $P_{\B, k+1}$, $P_{\S, k+1}$ and $P_{R, k+1}$, as given by a fixed fee schedule, and decide whether or not to subscribe to the platform for the next epoch.\looseness=-1 
	%
	\item The reward $r_k \sim \R(x_k, a_k)$ of the platform for epoch $k$ includes both the referral fees from epoch $k$ and the subscription fees collected from buyers and sellers, reflecting the decision in regard to whether or not to join the platform for epoch $k+1$.
	%
	This avoids delayed reward for the  platform, as the registration decisions, and thus registration fees for the next epoch, are influenced by the platform's matching strategy (i.e., action $a_k$) during  epoch $k$. 
	%
	%
	\item The platform's observability of information in the state follows the same as for the fee-setting POMDP. 
\end{itemize}

\section{Deferred Materials for Section~\ref{sec:experiments}}
\label{app:experiments}

\begin{table}[H]
	\centering
	\begin{tabular}{ll}
		\textbf{Parameters} & \textbf{Value}\\
		\hline\hline
		Observation size & 282\\
		Buyer/seller registration max & 10\\
		Seller referral max & 1\\
		Buyer/seller registration discretization & 0.2\\
		Seller referral discretization & 0.1\\
		Matching threshold discretization & 0.1\\
		Number of matching actions & 21\\
		\hline
		A2C algorithm &\\
		Optimizer & Adam\\
		Learning rate & 0.0001\\
		Batch size (fee-setting policy) & 4\\
		Batch size (matching policy) & 16\\
		Reward discount factor & 0.99\\
		Entropy weight & 0.01\\
		Training episodes & 30000\\
		Number of epochs per episode & 12\\
		Linear layer before LSTM & 256\\
		LSTM cell size & 128\\
		Linear layers for policy and value networks & 128\\
		\hline
		Bayesian Optimization &  \\
		Rounds & 64\\
		Initial points & 10\\
		Iterations in round & 50 \\
		\hline
		\hline
	\end{tabular}
	\vspace{2ex}
	\caption{Hyperparameters adopted for platform policy training.
	\label{table:hyperparam}}
\end{table}

\subsection{Experimental Results for Uniform and Two-Core Market Structures}
\label{sec:apx-other_latents}
\begin{figure}[H]
	\centering
	\begin{subfigure}{\columnwidth}	
		\centering
		\includegraphics[width=0.9\columnwidth]{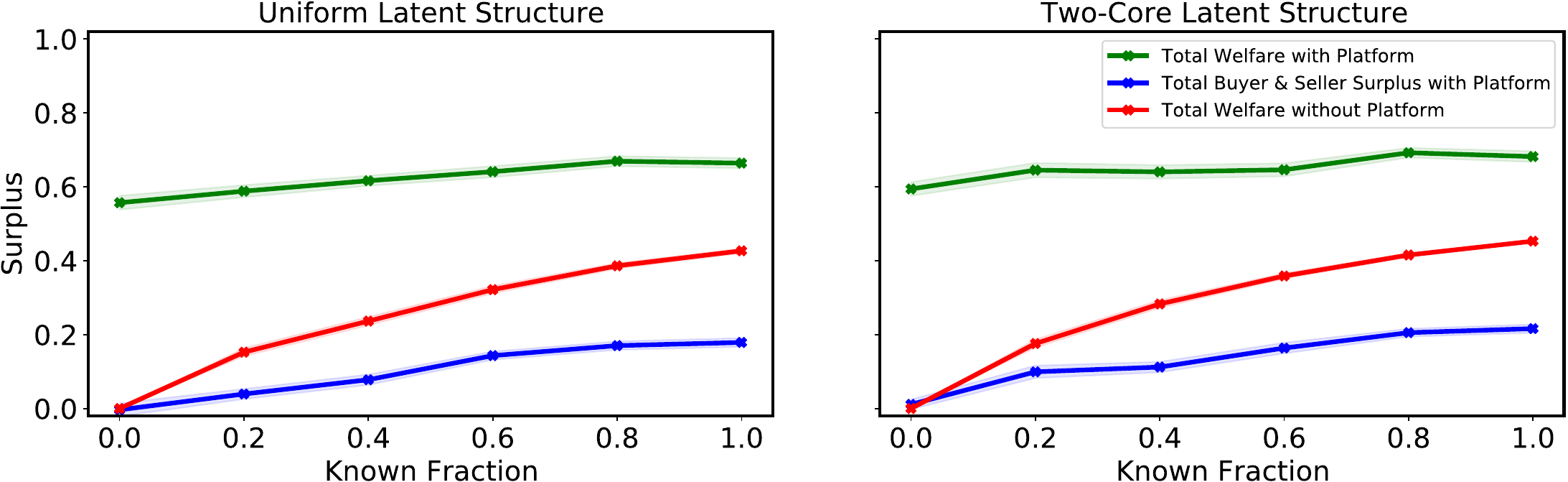}
		\caption{Varying knowledge level about sellers $\rho$, with  world transaction friction $\mu = 0.6$.}
	\end{subfigure} \\
	\vspace{0.05\columnwidth}
	\begin{subfigure}{\columnwidth}	
		\centering
		\includegraphics[width=0.9\columnwidth]{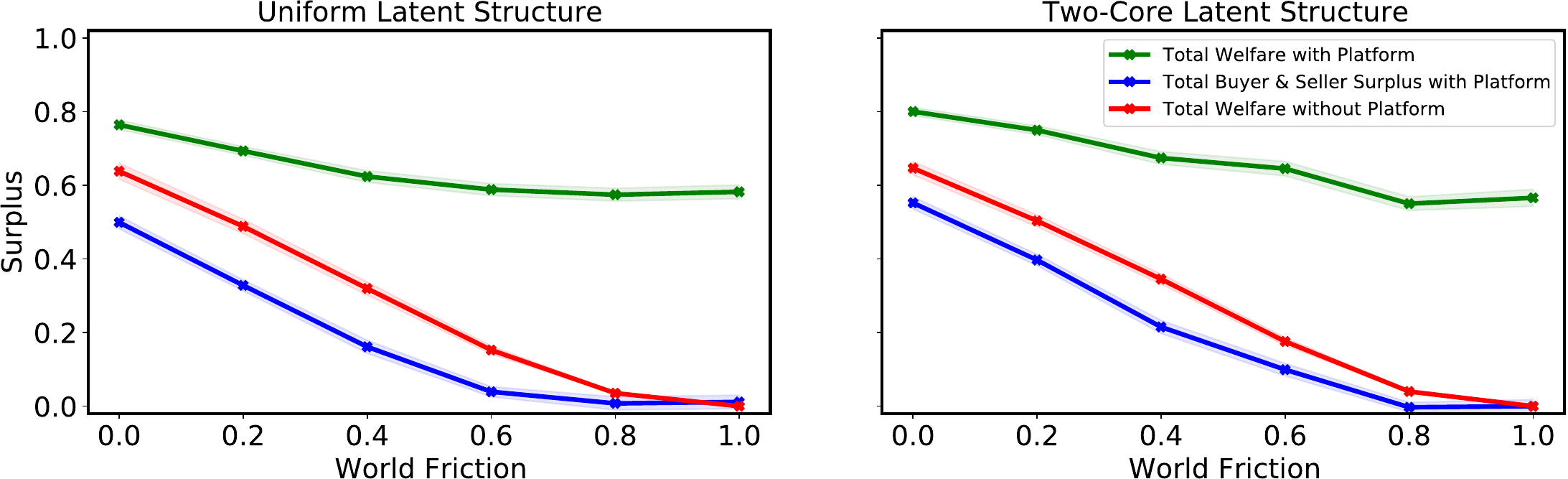}
		\caption{Varying  the world transaction friction $\mu$, with  knowledge level about sellers $\rho = 0.2$.}
	\end{subfigure}
	\caption{Total welfare and buyer and seller surplus, achieved in environments that vary in knowledge level and world transaction friction, with and without a platform under the \textit{uniform} and \textit{two-core} market structures. Results presented are the average of a hundred runs for each environment. 
		\label{fig:platform_value_latents}}
\end{figure}

Supplementing Figure~\ref{fig:platform_value_}, Figure~\ref{fig:platform_value_latents} shows the total welfare, as well as buyer and seller surplus, achieved in environments with and without a platform under the Uniform and Two-Core market structures.
As in the previously-seen graph for the Core-and-Niche structure, we normalize welfare and surplus by the total welfare achieved in an ideal world where buyers have complete knowledge about sellers and there is no world transaction friction.
\vspace{2ex}

In the sequel, we further present our main results of the \textit{uniform} and \textit{two-core} market structures, analyzing in the presence of market shocks, a no-intervention case as well as three different kinds of interventions. 
As shown in Figure~\ref{figure:all_obj_results_uniform} and Figure~\ref{figure:all_obj_results_two_core} below, the headline results remain qualitatively consistent to those of Core-and-Niche markets: taxation policies and caps imposed on some subset of fees are largely ineffective, whereas fixing fees to those that a self-interested platform chooses in an environment \textit{without} shocks while allowing the platform to continue to adapt its matching policy may successfully incentivize a revenue-maximizing platform to promote the seller diversity, efficiency, and resilience of the overall economy.
%

\begin{figure}[t]
	\centering
	\begin{subfigure}{\columnwidth}	
		\centering
		\includegraphics[width=\columnwidth]{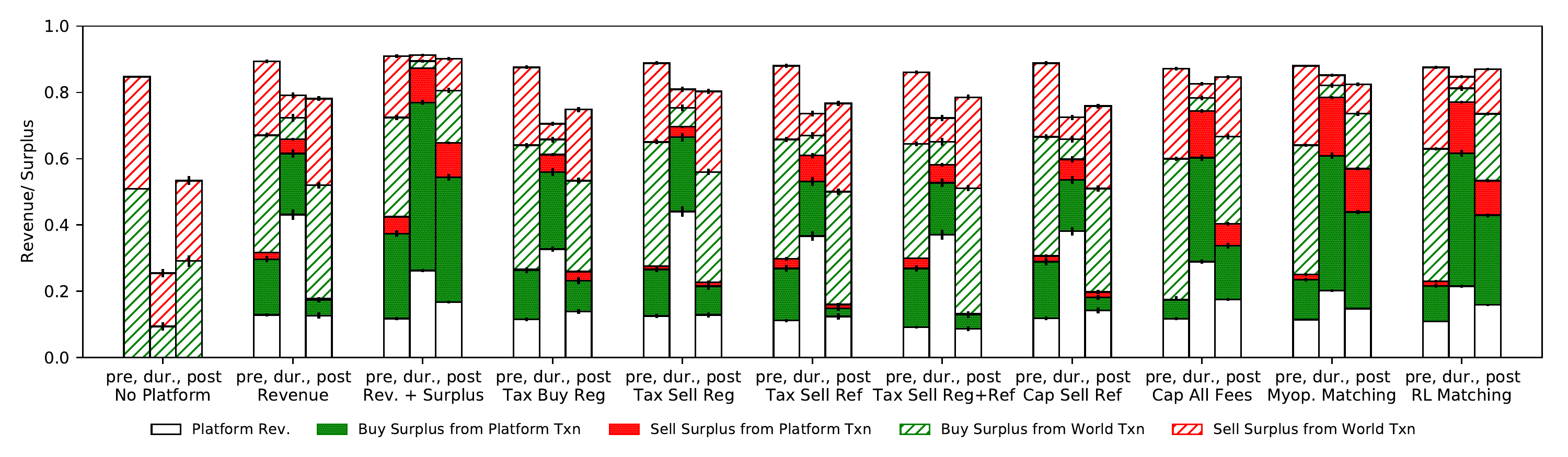}
		\caption{Welfare decomposition achieved by different learned fee-setting policies in each shock stage.}
		\vspace{2ex}
	\end{subfigure}
	\begin{subfigure}{\columnwidth}	
		\centering
		\includegraphics[width=\columnwidth]{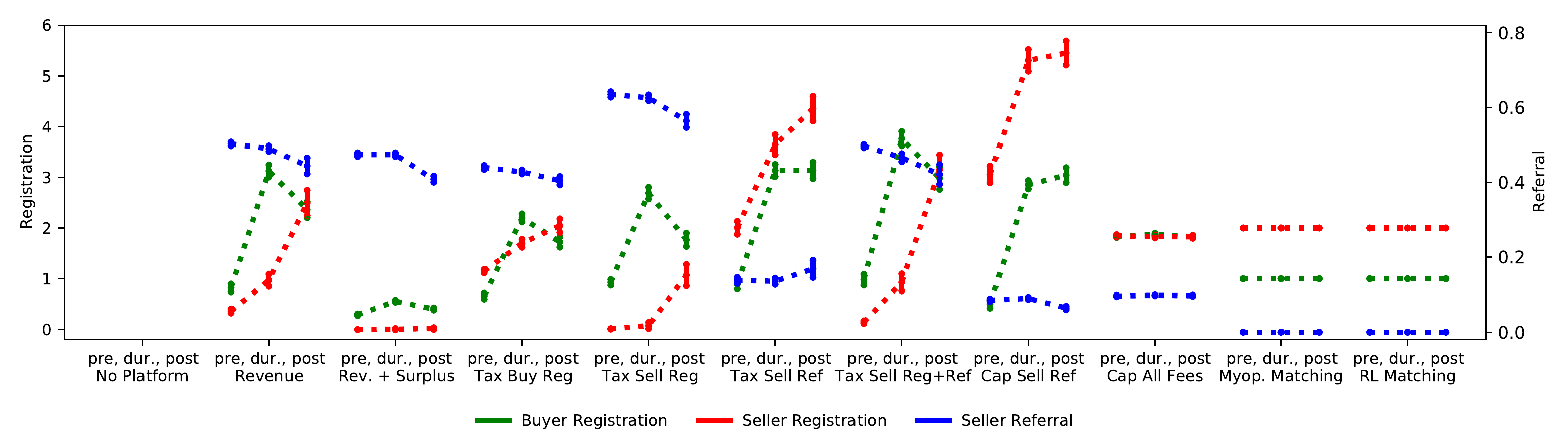}
		\caption{Platform fees set for each shock stage.}
		\vspace{2ex}
	\end{subfigure}
	
	\begin{subfigure}{\columnwidth}	
		\centering
		\includegraphics[width=\columnwidth]{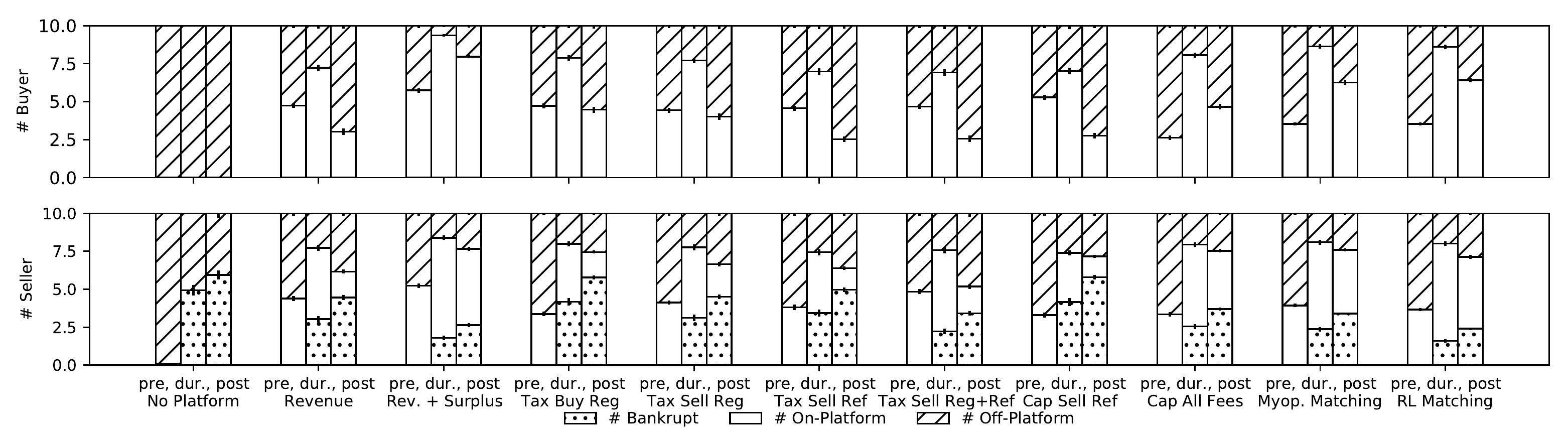}
		\caption{The number of on- or off-platform agents and bankrupt sellers induced by different regulatory interventions.}
	\end{subfigure}
	
	\caption{The welfare decomposition, platform fees, and agent states induced by different design objectives and considerations coming from regulation under the \textit{uniform} market structure.}
 
	\label{figure:all_obj_results_uniform}
\end{figure}

\begin{figure}[t]
	\centering
	\begin{subfigure}{\columnwidth}	
		\centering
		\includegraphics[width=\columnwidth]{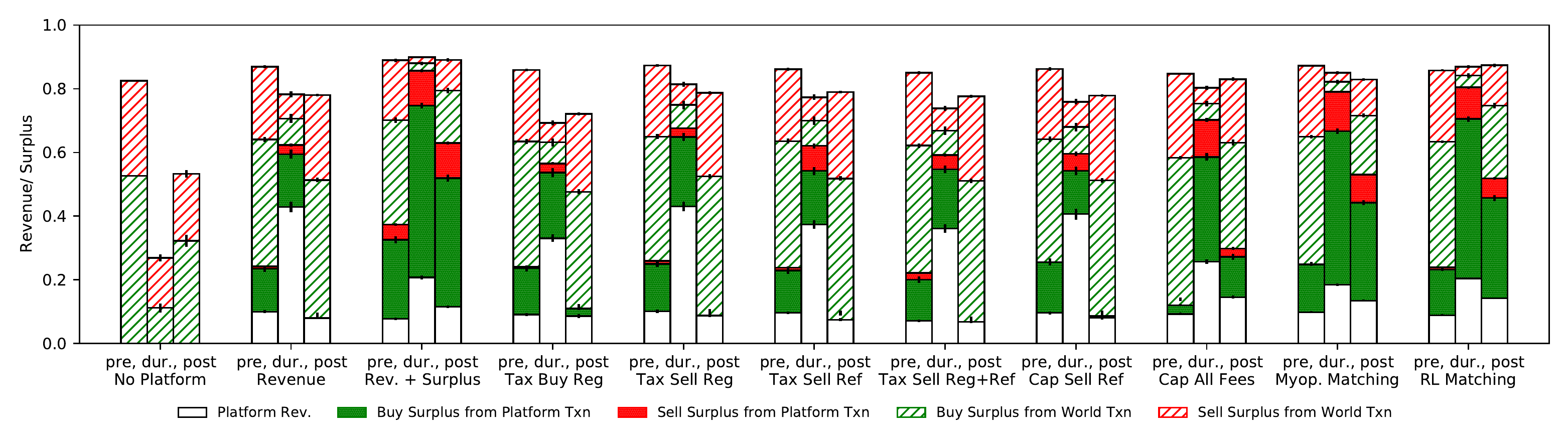}
		\caption{Welfare decomposition achieved by different learned fee-setting policies in each shock stage.}
		\vspace{2ex}
	\end{subfigure}
	\begin{subfigure}{\columnwidth}	
		\centering
		\includegraphics[width=\columnwidth]{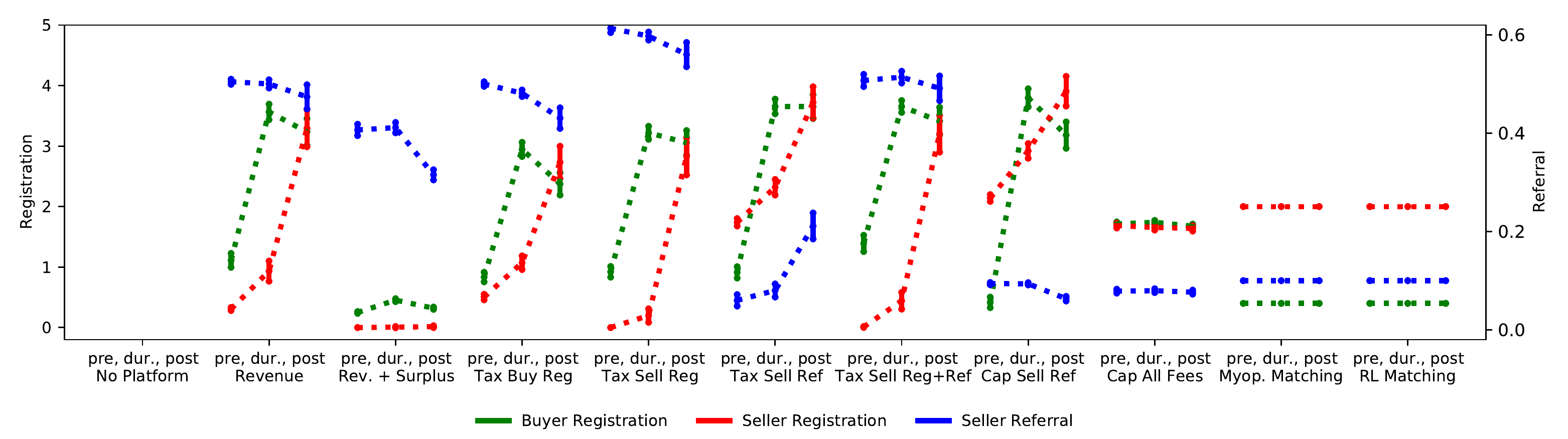}
		\caption{Platform fees set for each shock stage.}
		\vspace{2ex}
	\end{subfigure}
	
	\begin{subfigure}{\columnwidth}	
		\centering
		\includegraphics[width=\columnwidth]{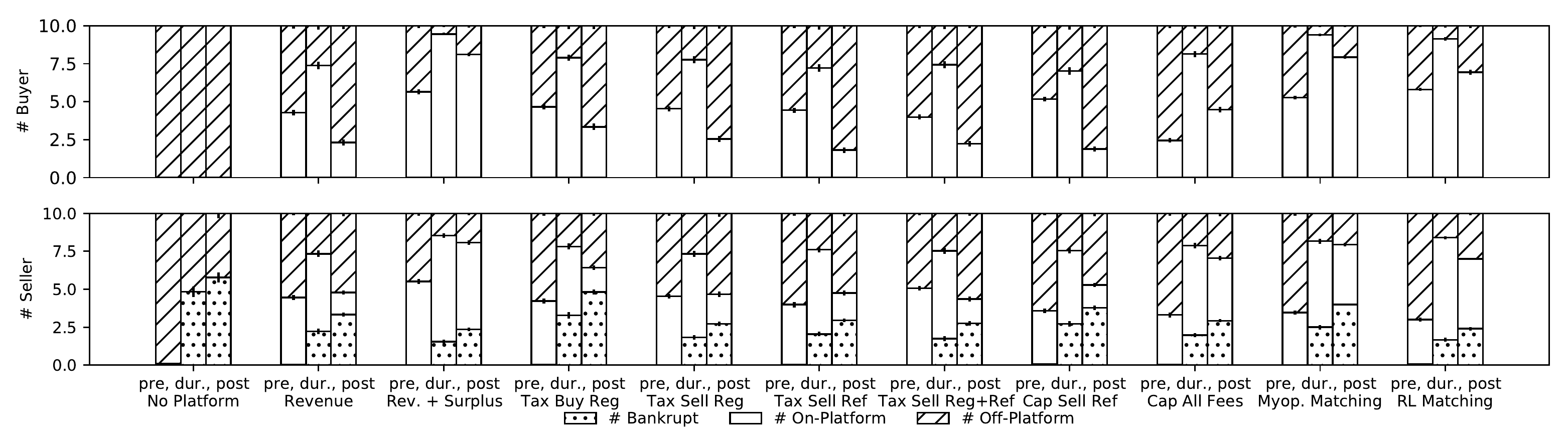}
		\caption{The number of on- or off-platform agents and bankrupt sellers induced by different regulatory interventions.}
	\end{subfigure}
	
	\caption{The welfare decomposition, platform fees, and agent states induced by different design objectives and considerations coming from regulation under the \textit{two-core} market structure.}
	\label{figure:all_obj_results_two_core}
\end{figure}

\end{document}